%% file: arxiv_mftrcost.tex
\documentclass[11pt]{article}

\usepackage{amsfonts,amsmath,amssymb,graphics}

\title{\bf Optimal multifactor trading under proportional transaction costs}

\newcommand{\LL}{\mathcal{L}}

\newcommand{\ex}{\mathbf{E}}

\newcommand{\V}{\mathbf{V}}
\newcommand{\util}{\mathcal{U}}
\newcommand{\vz}{\vec{z}}
\newcommand{\vn}{\vec{n}}
\newcommand{\vp}{\vec{p}}
\newcommand{\vZ}{\vec{Z}}

\newcommand{\notthis}[1]{}

\newcommand{\inv}{^{-1}}

\newcommand{\half}{\frac{1}{2}}

\newcommand{\cdl}{\,|\,}

\newcommand{\shalf}{{\textstyle\frac{1}{2}}}

\newcommand{\pderiv}[2]{\frac{\partial{#1}}{\partial{#2}}}

\newcommand{\pmderiv}[3]{\frac{\partial^2{#1}}{\partial{#2}\partial{#3}}}
\newcommand{\pdderiv}[2]{\frac{\partial^2{#1}}{\partial{#2}^2}}

\newcommand{\money}{G}

\newcommand{\tcmbuy}{\varepsilon_+}
\newcommand{\tcmsell}{\varepsilon_-}
\newcommand{\Corr}{\mathrm{Corr}}

\newcommand{\Time}{\textrm{time}}
\newcommand{\invar}{W}

\begin{document}
\author{Richard J. Martin}
\maketitle

\begin{abstract}

Proportional transaction costs present difficult theoretical problems in trading algorithm design, on account of their lack of analytical tractability. The author derives a solution of DT-NT-DT form for an arbitrary model in which the the traded asset has diffusive dynamics described by one or more stochastic risk factors. The width of the NT zone is found to be, as expected, proportional to the cube root of the transaction cost. It is also proportional to the $\frac{2}{3}$ power of the volatility of the target position, thereby causing a faster trading strategy to be buffered more than a slower one. The displacement of the middle of the buffer from the costfree position is found to be proportional to the square of the width, and hence to the $\frac{2}{3}$ power of the transaction cost; the proportionality constant depends on the expected short-term change in position.

\end{abstract}


\section{Introduction}

In this paper we consider the effect of proportional\footnote{The term `linear' costs often refers to the presence of fixed per-ticket cost and also a proportional part generated by a bid-offer independent of the trade size. As we are not considering a fixed part, we use the term `proportional', whereas in \cite{Martin11a} the term `linear' was used for the same thing.} transaction costs on the trading of an arbitrary `synthetic' asset that follows a diffusion process whose drift and volatility terms are driven by stochastic factors. The tradable asset is $X_t$ which in real probability measure evolves according to
\begin{equation}
dX_t = \mu_X(\vZ_t)\, dt + \sigma_X(\vZ_t)\,dW_{0,t}
\label{eq:model1}
\end{equation}
in which $\mu_X$ and $\sigma_X$, which describe the drift and volatility\footnote{Absolute, rather than lognormal, volatility.} of $X_t$, are smooth\footnote{Continuous first derivative; though mild singularities, such as that of $|z|$ at $z=0$, are not in fact a problem. We also need $\sigma_X>0$.} functions, and $\vec{Z_t}$ a vector of factors each component of which follows its own ergodic process:
\begin{equation}
dZ_{i,t} = \mu_{Z_i}(\vZ_t)\, dt + \sigma_{Z_i}(\vZ_t)\,dW_{i,t}.
\label{eq:model2}
\end{equation}
Call the position that one would trade to in the absence of costs the `target position'. In the presence of proportional transaction costs, one cannot simply follow the optimal costfree strategy, as to do so would lose money at an infinite rate.  Instead, a `buffer' must be drawn around the target position, defining a no-trade (NT) zone in which the position is held unchanged and on either side a discrete-trade (DT) zone in which one trades immediately to the edge of the NT zone\footnote{Note incidentally that the DT-NT-DT form of solution is not needed in the effectively unrelated, and simpler, problem of purely quadratic costs, i.e.\ total cost $\propto (\mbox{trade size})^2$ with no proportional or fixed terms. In that setup one instead trades continuously torwards a target position. As pointed out by Garleanu and Pedersen \cite{Garleanu09}, with purely quadratic costs, when one determines what weight should be attached to all the various signals at one's disposal, the fast signals receive only a low weight as the slowness of trading causes them to be smoothed out.}. The effect of the NT zone is to prevent the strategy trading backwards and forwards in small  amounts: typically the action is to execute a succession of small trades in one direction, then wait, then reverse. One must find the optimal buffer width. Too narrow, and one loses too much in costs by overtrading; too wide, and trading is performed so rarely that no revenue is generated.

In a recent paper \cite{Martin11a}, a very specific one-dimensional example of (\ref{eq:model1}) is considered, in which $Z_1\equiv X$, and the exact shape of the optimal buffer (i.e.\ where its ends should be, as a function of $X_t$) is derived in terms of some functions relating to $\mu_X$, $\sigma_X$. Analysis of this solution by Taylor series shows that in the limit of small costs the buffer width is, at leading order, proportional to the cube root of the cost $\varepsilon$  of trading one unit of asset, and the constant of proportionality is given (see eq.~\ref{eq:soln-0f}). It is also shown that the middle of the buffer is not quite at the target position (see eq.~\ref{eq:displ-0f}), but this is a smaller effect.
 In this paper, we extend to the multifactor case (\ref{eq:model1}). We first present an heuristic argument for the results. In the Appendix we give a full derivation the key idea to which is the reduction of the problem to a one-dimensional system of the same type as \cite{Martin11a}. We thereby obtain a similar optimal trading rule, given by (\ref{eq:soln-mf-1},\ref{eq:displ}), and demonstrate via model examples using synthesised and real data that it works quite well.

\subsection*{Background}

We can think of $X_t$ being a futures price, a combination of these, or more generally some kind of swap PV, and so it is allowed to be negative; there is no term corresponding to the risk-free interest rate in its drift, because it is not the PV of a cash asset\footnote{So in the risk-neutral world it would have zero drift.}. Thus for example rather than trading stocks and cash bonds, we are taking risk through their synthetic analogues, namely stock futures, bond futures, interest rate swaps and CDS. (For a general background to this setup, and to transaction costs in general, the reader is referred to \cite{Martin11a}.)

The factors represented by $\vZ_t$ may be exogenous (specified by the modeller, as for example macroeconomic ones) or endogenous (intrinsic to the time series of $X_t$, as for example momentum), and for these purposes it does not matter which, but there are some subtleties of interpretation that are more conveniently discussed later.

The ergodicity assumption is a subtle one: although there is no explicit requirement for the factors to be mean-reverting, it is usually necessary when defining a time series model for a realisation-average, i.e.\ an average over a probability distribution, to be interchanged with a time-average. This implicitly happens when the model is traded: the effect of making decisions that are `on average correct' (=realisation average) becomes apparent as time progresses. It also occurs when a model is back-tested, as one hopes that a reasonable coverage of scenarios (realisation average) has been obtained in the course of history (time average). The universally-employed technique of standardising a factor, that is, subtraction of its unconditional mean followed by division by its unconditional standard deviation, naturally causes it to become mean-reverting. There is \emph{no} implication in this paper that $X_t$ itself is mean-reverting, as the functions $\mu_X$, $\sigma_X$ are arbitrary.



\subsection*{Optimality, objective, and utility}

Our construction is `optimal', in the sense of maximising the value function\footnote{The treatment here is incidentally along the same lines as \cite{Kallsen99}.}
\begin{equation}
V_t =  \ex_t \left[ \int_{s=t}^\infty  e^{-r(s-t)} \util(\theta_s  \, dX_s) \right]
\label{eq:valfn}
\end{equation}
where $\theta_t$ is the position\footnote{Number of lots, for futures contracts; notional, for OTC contracts such as swaps} in $X_t$ at time $t$ and $\util$ denotes the utility of instantaneous \emph{changes in} P\&L. As the dynamics are diffusive, without jumps, we only require $\util$ and its first two derivatives at the origin. We stipulate $\util(0)=0$, $\util'(0)=1$, $\util''(0)=-1/\money$, so that $\money$, which is fixed, is a measure of risk appetite and has units of money (because $\util$ does, in our formulation).
Then we can recast (\ref{eq:valfn}) as
\begin{equation}
V_t =  \ex_t \left[ \int_{s=t}^\infty  e^{-r(s-t)} \bigg(\theta_s \mu_{X_s} - \frac{\theta_s^2 \sigma^2_{X_s}}{2G} \bigg) \, ds\right]
\label{eq:valfnmv}
\end{equation}
which makes it clear that the objective function rewards upward drift, and penalises volatility, in P\&L.
We seek an optimal solution $\theta_t$ that depends only on the factors at time $t$ and the current position $\theta_t^-$, writing $\theta_t=g(\vZ_t,\theta_t^-)$.

In the absence of transaction costs the optimal position is simply
\begin{equation}
\hat{\theta}_t \equiv \hat{g}_0(\vZ_t) = \frac{\mu_X(\vZ_t)}{\sigma_X(\vZ_t)^2} \money
\label{eq:theta_ntc}
\end{equation}
so that one trades to this target position without reference to the current position.
(The $_0$ in $\hat{g}_0$ indicates the transaction-free solution, and the $\hat{ }$ denotes optimality.) This is of the familiar form ``expected return $\div$ variance, $\times$ gearing factor'' (see also the derivation in \cite[\S14]{Bjork98}).

\subsection*{Interpretation of the gearing factor $\money$}

The gearing factor expresses the utility of risk, and can be understood by appeal to volatility of P\&L.
The expected return in a time interval $dt$ is $\theta_t\mu_X\, dt$ and the variance of the P\&L is $\theta_t^2 \sigma_X^2\, dt$. By (\ref{eq:theta_ntc}), the mean P\&L, the variance of the P\&L, and the Sharpe ratio for a time period $[0,T]$ are
\begin{equation}
\ex \left[ \int_0^T \frac{\mu_{X_t}^2 \money}{\sigma^2_{X_t}} \, dt\right] ; \qquad
\ex \left[ \int_0^T \frac{\mu_{X_t}^2 \money^2}{\sigma^2_{X_t}} \, dt \right] ; \qquad
\mathfrak{r}_T = \left( \ex\left[ \int_0^T \frac{\mu_{X_t}^2}{\sigma^2_{X_t}} \, dt  \right]\right)^{1/2}.
\label{eq:mvsr}
\end{equation}
So the standard deviation (stdev) of the P\&L is simply $\mathfrak{r}_T \money$. The interpretation of $\money$ is now clear: it is the desired stdev of P\&L, per unit Sharpe. Suppose that a strategy for trading a particular asset has an (estimated) annual  Sharpe ratio of 0.8 and the desired annual P\&L stdev is \$60M. Then the gearing is set to $\money=\,$\$75M. 

Another interpretation of $\money$ is as a Lagrange multiplier. Consider the usual optimisation: Maximise return subject to risk $\le$ some value $Q$ say. If we understand risk to be quadratic variation then the Lagrangian is
\[
\theta \ex[dX_t] + \lambda ( \theta^2 \V[dX_t] - Q\, dt)
\]
and $\lambda=-1/2G$.

In practice, in a fund, gearing is done as follows. First, optimise and provide a backtest of the optimised strategy using, for example, $\money=\,$\$1M. This provides an estimate of the historical Sharpe ratio and P\&L volatility. Other strategies are treated similarly. At a higher level, the portfolio manager uses this information, and the correlations between strategies, to determine an `allocation' to each, or, in effect, a level of P\&L volatility for each to run: one then scales one's $\money$ to suit. Should all strategies have the same $\money$, then the risk being run on each is directly proportional to its Sharpe. This assumes, as we are doing here, that transaction costs are proportional and there is no degradation in Sharpe as more risk is run. 


\subsection*{Summary of \cite{Martin11a}}

The work in \cite{Martin11a}, and here, is a departure from current literature such as \cite{Davis90,Shreve94,Whalley97,Zakamouline06} in a few respects. First, we optimise an objective function that has an infinite horizon, not a finite one, and is dependent not on terminal utility but on quadratic variation (in a more general setting, utility of P\&L variation). Hence the upper limit in (\ref{eq:valfn},\ref{eq:valfnmv}) is infinite, and the value function satisfies an ordinary differential equation (ODE) rather than a PDE. This enables the value function to be written down easily as a function of the DT-NT-DT geometry, and then optimised.

Another important distinction is that risk is taken synthetically here. Whereas in the Merton problem there is a riskless asset and a risky asset (e.g.\ a stock), here we take risk through a futures position, trading one asset only. This means that we have different boundary conditions; in fact we theoretically permit unlimited loss but in practice the probability of this is too small to be significant.

The main result of \cite{Martin11a} is that the optimal NT boundary can be determined exactly, more or less in closed form through a pair of coupled nonlinear (but not differential) equations. Using Taylor series expansion, one can study this solution in the limit of small transaction costs\footnote{This is rigorously justified using the (Analytic) Implicit Function Theorem. The algebra is lengthy, but not conceptually difficult, and the problem is much more tractable than having to deal with the PDEs that arise in other authors' work. One does not have to guess that an expansion in powers of $\varepsilon^{1/3}$ is necessary: it is a immediate consequence of the Taylor series expansion. More details are in the Appendix and in the full version of \cite{Martin11a}.}, and  finds that the half-width of the NT zone (buffer) is, expressed as an amount of asset to be traded\footnote{We can think of the buffer in `position space', as in the present description, or in `factor space', i.e.\ the amount the factor(s) must move to trigger a trade. We shall in due course need both ideas.}\,\footnote{Result for the halfwidth is corroborated, using entirely different methods, in the case where $X$ follows an Ornstein-Uhlenbeck process, by Bouchaud \& co-workers in \cite{Lataillade12}.},
\begin{equation}
\delta \theta  = \left(\frac{3\varepsilon \money |\hat{g}_0'(X_t)|^2}{2}\right)^{1/3}+ O(\varepsilon),
\label{eq:soln-0f}
\end{equation} 
and the displacement of the buffer from the costfree position is
\begin{equation}
\mathrm{d} \theta  = -\theta \left(\frac{2\varepsilon^2 |\hat{g}_0'(X_t)|}{3\money }\right)^{1/3}+ O(\varepsilon^{4/3}),
\label{eq:displ-0f}
\end{equation} 
where $\varepsilon$ is the cost of buying or selling one lot of $X$. (N.B. Do not confuse $\delta\theta$ with $\mathrm{d}\theta$.)
Both expressions are directly proportional to the gearing $\money$, as expected, because $\hat{g}_0$ contains a factor of $\money$ also. The cube-root dependence of $\delta\theta$ on cost is a well-known result. Higher powers go up in steps of $\varepsilon^{2/3}$. In single-period models, or situations in which one trades into a position and then holds it indefinitely, there is no $O(\varepsilon^{1/3})$ term and the behaviour is $O(\varepsilon)$: thus the $O(\varepsilon^{1/3})$ behaviour arises from continuous trading. Informally, the constant of proportionality deals with the  variability of the target position: the more variable, the wider the buffer needs to be.

As our setup is not the same as that of other authors, the reader needs to exercise caution about comparing the results shown here with others: whereas there are clear similarities, such as the $\varepsilon^{1/3}$ dependence, and (less importantly) the $\frac{3}{2}$ factor, the remaining terms are not quite the same. The interpretation of $|\hat{g}_0'(X_t)|^2$ as a trading speed is important and it is absent from the basic Merton problem. In the Merton problem one rebalances stock and riskless asset so as to achieve a constant proportion, which we regard to be of little practical relevance. In this paper, rebalancing occurs because the factors, and hence the view of the profitability of the asset, and hence the desired trading position, change: in our view this is a much more satisfactory model of the way that investment management works.

\subsection*{Outline and summary of this paper}

The derivation is given in the Appendix and consists in reworking the derivations of \cite{Martin11a}. The result is that the halfwidth is\footnote{$\V_t$ denotes the variance, conditional on information known at time $t$.}
\begin{equation}
\delta \theta  \sim \left(\frac{3\varepsilon \money \hat{\Gamma}_0^2}{2}\right)^{1/3},
\qquad \hat{\Gamma}_0^2 = \frac{\V_t[d\hat{g}_0(\vZ_t)]}{\V_t[dX_t]}
\label{eq:soln-mf-1}
\end{equation}
and the displacement is
\begin{equation}
\mathrm{d}\theta 
\sim
\frac{\ex_t[d\hat{g}_0(\vZ_t)]}{\V_t[dX_t]} \left(\frac{2\varepsilon^2G^2}{3\hat{\Gamma}_0^2} \right)^{1/3} .
\label{eq:displ}
\end{equation}
We now discuss these results in turn and present some heuristic justifications for them.

For the half-width, we have reinterpreted $\hat{\Gamma}_0^2$ in (\ref{eq:soln-0f}) as the ratio of the variation of the target position to the variation of the tradable asset:
\begin{equation}
|\hat{g}_0'(X_t) |^2
  \rightsquigarrow 
\frac{\V_t[d\hat{g}_0(\vZ_t)]}{\V_t[dX_t]}
= \frac{ \sigma^2_{\hat{g}_0} }{ \sigma^2_X }
.
\label{eq:dgdxnew}
\end{equation}
In effect, what we have done is reinterpret $|d\hat{g}_0/dX|^2$ as $(d\hat{g}_0)^2/(dX)^2$ and taken expectations.
Clearly something had to be done about the $|\hat{g}_0'(X_t)|^2$ term in (\ref{eq:soln-0f}), for whereas in \cite{Martin11a} the target position depended only on the tradable, here it depends on the factors, so $\hat{g}_0'(X_t)$ is not meaningful in this context. In the case of \cite{Martin11a}, the two expressions in (\ref{eq:dgdxnew}) coincide. However, (\ref{eq:displ}) cannot be obtained by simply replacing $|\hat{g}_0'(X_t)|$. Note also that the displacement is $O(\varepsilon^{2/3})$; it is therefore of less importance when costs are small.

A few obvious points can be made about (\ref{eq:soln-mf-1},\ref{eq:displ}). As expected the RHS is directly proportional to $\money$ because $\hat{g}_0$ is, so the effect of $\money$ is not interesting (and does not affect, for example, the trading speed of the model). It can also be seen that the buffer is wider when the attempted trading speed is higher, i.e.\ $\hat{\Gamma}_0^2$ is higher. The dependence on the volatility $\sigma_X$ of the traded asset is more subtle. There is an explicit factor of $\sigma_X$ in the expression for $\hat{\Gamma}_0^2$, and other factors in the expression for $\hat{g}_0$. The effect is to reduce the buffer width \emph{and also the target position} as $\sigma_X$ increases; informally, the buffer width remains roughly the same as a proportion of the `typical' target position, an issue that will become clearer when we make explicit calculations later on.

The results can be justified by scaling arguments\footnote{If one is prepared to take the numerical factors on trust.} 
: a fuller discussion is in \cite{Rogers04}.
The reduction in expected utility consequent on trading to the edge of a buffer of width $\delta\theta$, rather than to the target position itself, is $\propto \delta\theta^2$, as the utility is quadratic at its maximum. More precisely let $\dot{U}$ denote the rate of accumulation of expected utility in (\ref{eq:valfn},\ref{eq:valfnmv}), i.e.\
\[
\dot{U}(\vz,\theta) = \frac{1}{dt} \ex_t\big[\util(\theta \, dX_t) \cdl \vZ_t=\vz \big]
= \mu_X(\vz)\theta  - \frac{\sigma_X(\vz)^2\theta^2}{2\money};
\]
then the reduction is $\half \,\delta\theta^2 \partial_2^2 \dot{U}$ which is equal to $-\sigma_X^2\,\delta \theta^2/2\money$.
The time taken to exit a buffer of width $\delta\theta$ scales as $\delta \theta^2$ divided by the square of the volatility of the target position, i.e.\ $\delta\theta^2\big/\sigma^2_{\hat{g}_0}$. The cost associated with this is $\varepsilon\,\delta\theta$. Adding the two parts together gives the total utility loss rate through suboptimal positioning and through explicit payment of transaction costs.
We are therefore to maximise
\[
- \frac{\sigma_X^2}{2\money} \delta \theta^2 - \frac{\varepsilon \,\delta\theta}{\delta\theta^2\big/ k\sigma^2_{\hat{g}_0}}
\]
for some positive constant $k$, from which it is easily seen that the maximum occurs when $\delta \theta$ is given by the expression in (\ref{eq:soln-mf-1}). That $k=\frac{3}{2}$ comes from comparison with (\ref{eq:soln-0f}). 

An extension of this argument gives a result for the displacement, as follows. Suppose that the drift $\mu_{\hat{g}_0}$ in target position is positive, and that we currently have too high a position on, so that $\theta_t^-$ exceeds the target position. If we sell now, it is likely that we will be undoing the trade as the target position drifts back up, thereby resulting in an extra dose of transaction cost. The extra cost is $\varepsilon\, \mathrm{d} \theta$ and the time taken to drift towards the edge of the buffer is of order $\delta \theta /\mu_{\hat{g}_0}$.  The edges of the buffer are at $(\delta \theta \pm \mathrm{d}\theta)^2$ of which the average is $\delta \theta^2 + \mathrm{d}\theta^2$ so we insert this into the first term in the objective function (the penalty for having a suboptimal position on).
The revised objective function is
\[
- \frac{\sigma_X^2}{2\money} (\delta \theta^2 + \mathrm{d}\theta^2) - \frac{\varepsilon \,\delta\theta}{\delta\theta^2\big/ k\sigma^2_{\hat{g}_0}} + \frac{\varepsilon \, \mathrm{d}\theta}{\delta\theta\big/k'\mu_{\hat{g}_0}},
\]
which gives, on maximising w.r.t.\ $\mathrm{d}\theta$,
\[
\frac{\sigma_X^2}{\money} \mathrm{d}\theta = \frac{\varepsilon}{\delta\theta\big/ k'\mu_{\hat{g}_0}}
\]
for some other positive constant $k'$. Consistency with (\ref{eq:displ-0f}) requires $k'=1$ and we deduce
\begin{equation}
\delta\theta \, \mathrm{d}\theta = \frac{\mu_{\hat{g}_0}}{\sigma_X^2} \varepsilon  \money
\end{equation}
which in conjunction with (\ref{eq:soln-mf-1}) gives (\ref{eq:displ}).
The revision to the objective function does not vitiate the result we just derived for $\delta \theta$ (at leading order).

The neatness of (\ref{eq:soln-mf-1}) is mathematically appealing and the formula is very easy to implement; an exact solution is very difficult to derive, though it was done in \cite{Martin11a} for one factor $Z_1\equiv X$.
We regard (\ref{eq:displ}) as less important: it is a smaller effect and the position-drift term is rather harder to estimate, so in our numerical work we centre the buffer around the target position, implicitly assuming that $\mathrm{d}\theta=0$.

The purpose of the next sections is to demonstrate the buffering rule (\ref{eq:soln-mf-1}) using a variety of models of differing complexity to assess its empirical validity. We first do this in a `controlled experiment' using synthesised data from a model in which all parameters are known exactly, and find that it works well. Then we fit a momentum model to some real financial data, on which the underlying model and parameterisation can only be estimated, and find that it still works reasonably well.


\section{Examples using synthesised data}

\subsection*{Objective}

We first consider some models with synthesised data that will incorporate effects seen in real time series. There are advantages to this: we can separate out different effects in a controlled experiment, and we can simulate as much data as we please. We could simulate the various Brownian motions driving the factors, compute the positions and P\&L net of transaction costs and hence the (discounted) utility, then repeat many times and average so as to obtain an estimate of the value function. However, a useful short-cut is to exploit the ergodicity of the models and replace a realisation-average with a time-average.
Thus we simulate instead only one trajectory, for a long time period (and the discount factor can safely be ignored\footnote{It is absent from the approximate solution for the optimal buffer width and displacement.}). Explicitly the following quantity, which we call the \emph{empirical value function}, is being maximised:
\[
V_\mathrm{emp} =  \sum_{i=0}^{N-1} \util \big(\theta_{t_i} (X_{t_{i+1}}-X_{t_i})\big) - \varepsilon \big|\theta_{t_{i+1}}-\theta_{t_i}\big|.
\]
In discrete time it will be necessary to specify an explicit utility function, rather than just its first two derivatives at the origin. We have used $\util(x)=(1-e^{-x/\money})\money$ throughout.
By the (empirical) account curve, we mean the time series of integrated P\&L after costs, i.e.
\[
\sum_{i=0}^{n-1} \theta_{t_i} (X_{t_{i+1}}-X_{t_i}) - \varepsilon \big|\theta_{t_{i+1}}-\theta_{t_i}\big|, \qquad 1 \le n \le N.
\]
Analogously with real data we do exactly the same thing. Of course, both $V_\mathrm{emp}$ and the account curve are random variables, but the idea is that the implicit randomness is attenuated by observing for long enough time. Also, plotting a single time series and account curve makes for more convenient interpretation and illustration.

We wish to demonstrate whether the approximate buffering rule (\ref{eq:soln-mf-1}) is optimal, but this presents us with a difficulty, as in principle we must test against all other buffering schemes. Another difficulty is that the buffer width is in general varying, so we cannot simply plot the buffer width against the empirical value function, as there is no unique buffer width to plot. What we can do easily, though, is multiply $\delta\theta$ in eq.~(\ref{eq:soln-mf-1}) by some fixed amount $\lambda$; we then plot the \emph{time-average} of the buffer width on the horizontal axis, and  on the vertical axis the empirical value function. Repeating for different values of $\lambda$ causes a curve to be described, and we highlight the point corresponding to $\lambda=1$. Finally, we repeat for different transaction cost parameters to give a family of curves. 
Consider what the curve should look like, as a function of $\lambda$. If the buffer width is too small ($\lambda\to0$) then the value function will drop (in continuous time it would drop to $-\infty$): the drop will be severe if $\varepsilon$ is high. If the buffer is too wide ($\lambda\to\infty$) the NT zone will become so large that no trading takes place, and the value will tend to zero. At some intermediate point there should be a maximum; ideally the highlighted point $(\lambda=1)$ will be at the hump, indicating that no improvement can be made by scaling (\ref{eq:soln-mf-1}) up or down by a fixed amount (though it does not rule out the possibility that the buffer is suboptimal by virtue of being at some times too wide and at other times too narrow). However, if the costs are high enough, the value function will always be negative and there will be no hump: then the strategy is worthless, irrespective of how well `optimised' the buffer is\footnote{As an ansatz the reader might wish to plot $-\varepsilon w^{-1}+(1+w^2)^{-1/2}$ vs $w$, for $0<w<\infty$. For $\varepsilon<1$, the curve has a local maximum above zero at $\hat{w}$ obeying $\hat{w}=\varepsilon^{1/3}(1+\hat{w}^2)^{1/3}$. But for $\varepsilon\ge1$, it has none and it is always negative.}. Note that throughout we ignore the buffer displacement in our simulations, assuming it to be zero.

\subsection*{Linear model}

We say that the model is linear, if $\mu_X$ depends linearly on $\vZ$ and $\sigma_X$ is constant, with $\vZ$ a multivariate OU process. The simplest example is the one-factor model,
\begin{eqnarray}
dX_t &=& \beta\sigma Z_{1,t} \, dt + \sigma \, dW_{0,t} \label{eq:modlin} \\
dZ_{1,t} &=& -\kappa_1 Z_{1,t} \, dt + \sqrt{2 \kappa_1} \, dW_{1,t} \nonumber .
\end{eqnarray}
This is a simple momentum model, parameterised by $\beta$ (the strength of trending), $\sigma$ (the volatility of the traded instrument), $\kappa_1$ (the rate of switching of trend) and $\rho_{01}=\Corr(dW_{0,t},dW_{1,t})$, the correlation between changes in the factor and the tradable. One simulates $W_{0,t}$ and $W_{1,t}$ first and from that $Z_{1,t}$ and thence $X_t$. It is easily seen that
\[
\hat{g}_0(\vZ_t) = \frac{\beta Z_{1,t}\money}{\sigma} ; \qquad
\hat{\Gamma}_0^2 = \frac{2\beta^2\kappa_1G^2}{\sigma^4};
\]
hence
\[
\delta\theta \sim \left(\frac{3\varepsilon\kappa_1}{\sigma|\beta|}\right)^{1/3} \frac{\money |\beta|}{\sigma} 
= 
 \left(\frac{3\widehat{\varepsilon}\kappa_1}{|\beta|}\right)^{1/3} \langle \theta^2 \rangle^{1/2} 
\]
where $\widehat{\varepsilon}=\varepsilon/\sigma$ is the cost per unit volatility of the tradable and $\langle \theta^2 \rangle^{1/2} = \money\beta/\sigma$ is the r.m.s.\ position\footnote{Root mean square. Note $\langle Z_1 \rangle =0$ and $\langle Z_1^2 \rangle =1$. Do not confuse $\langle\theta^2\rangle^{1/2}$ with $\sigma_{\hat{g}_0}$ which denotes the \emph{volatility} of the position and pertains to \emph{changes} in position. We could also write $\langle \theta^2 \rangle = \V_\emptyset[\theta_t]$ i.e.\ the variance of $\theta_t$ given no information: by stationarity, these are the same.}.
Notice that $\hat{\Gamma}_0^2$ and hence $\delta\theta$ are constant, and $\rho_{01}$ does not play a part. Notice also that the buffer width and position are both inversely proportional to $\sigma$, provided one fixes $\hat{\varepsilon}$. (If the volatility of the underlying increases with $\varepsilon$ fixed, then the asset has actually become cheaper to trade and the buffer width drops as a fraction of the typical position.) Thus the only factors that link the buffer width to the r.m.s.\ target position are $\widehat{\varepsilon}^{1/3}$ and an extra quantity $\kappa_1/|\beta|$ that has dimensions $\Time^{-1/2}$; this is necessary for dimensional agreement (because $\widehat{\varepsilon}$ has dimensions $\Time^{1/2}$) and can be thought of as the trading speed, because the higher $\kappa_1$ is the more rapidly the momentum factor is changing direction. Finally, if $\beta\to0$ then the buffer width becomes large as a fraction of the r.m.s.\ position (not in absolute terms because the r.m.s.\ position reduces too): the explanation for this is that the asset price has in effect become less predictable, or that the trading signal is of lower quality: as expected, therefore, the NT zone becomes relatively wide and cuts down the amount of trading.

The displacement of the buffer is given by
\[
\mathrm{d}\theta \sim -\hat{\theta} \left(\frac{\hat{\varepsilon}^2\kappa_1^2}{3 \beta^2}\right)^{1/3}  
\]
thereby reducing the magnitude of the position, compared with the target position, by the indicated factor. In the context of the examples we are about to show, this is typically a few percent, which justifies our decision to ignore it.

Figure~\ref{fig:util_vs_buf}(a) shows results with $\kappa_1=0.02$, $\beta=0.2$, $\sigma=0.5$, $\rho_{01}=0$, for transaction cost $\varepsilon=$ 0.02, 0.05, 0.1, 0.2, 0.5. The gearing is fixed at $\money=1\$$, and the position $\theta_t$ and the buffer size are notional allocations to $X_t$ which here can be fractional\footnote{So for example a position $\theta=+2.157$ is permitted. When we do real contracts we will work with the proper contract sizes, so then one unit of $X$ will be `small', so one might in context be trading 2157 lots each of size 0.001.}. 
Note that the dimensions of $\kappa_1$, $\beta$, $\sigma$ are respectively $\Time^{-1}$, $\Time^{-1/2}$, $\$/\Time^{1/2}$, where $\Time$ is in the same units as the horizontal axis (which might be thought of as business days). The appearance of the graphs is as expected and the postulated rule (\ref{eq:soln-mf-1}) appears to be optimal. For low costs the impact of getting the buffer wrong is quite small, but for high costs it is much bigger.

\subsection*{Nonlinear coupling between tradable and factor}

We can replace $X$'s drift with a nonlinear function of the factor(s):
\begin{equation}
dX_t = \beta \sigma \, \gamma(Z_{1,t}) \, dt + \sigma \, dW_{0,t}.
\label{eq:modnonlin}
\end{equation}
Clearly
\[
\hat{g}_0(\vZ_t) = \frac{\beta \, \gamma(Z_{1,t})\money}{\sigma} ; \qquad
\hat{\Gamma}_0^2 = \frac{2\beta^2 \, \gamma'(Z_{1,t})^2\kappa_1G^2}{\sigma^4}.
\]

Figure~\ref{fig:util_vs_buf}(b) is a repeat of Figure~\ref{fig:util_vs_buf}(a), with $\gamma(z)=\tanh(2z)$ (this choice is arbitrary and inspired by the use of such sigmoidal functions in neural network theory, where they are referred to as \emph{activation functions}, see e.g.~\cite{\notthis{Cybenko89,}Haykin98}). Again the postulated rule (\ref{eq:soln-mf-1}) appears close to optimal.

\subsection*{Stochastic volatility}

We can introduce a volatility factor by exponentiating an OU process and using that in place of $\sigma_X$, as follows. With one return factor ($Z_1$) and one volatility factor ($Z_v$), one has:
\begin{eqnarray}
dX_t &=& \beta \sigma_t Z_{1,t} \, dt + \sigma_t \, dW_{0,t} \label{eq:modstochvol}  \\
\sigma_t &=& \overline{\sigma} \exp\big(\eta Z_{v,t}-\shalf \eta^2\big) \nonumber \\
dZ_{1,t} &=& -\kappa_1 Z_{1,t} \, dt + \sqrt{2 \kappa_1} \, dW_{1,t} \nonumber \\
dZ_{v,t} &=& -\kappa_v Z_{v,t} \, dt + \sqrt{2 \kappa_v} \, dW_{v,t} \nonumber
\end{eqnarray}
where in addition to the linear model $\eta$ is the relative amount by which the instantaneous volatility $\sigma_{t}$ varies, $\kappa_v$ is the rate at which it reverts, and $\rho_{1v}=\Corr(dW_{1,t},dW_{2,t})$ controls the extent to which volatility is correlated with asset price\footnote{In stock markets, for example, this is usually negative.}.
Then
\[
\hat{g}_0(\vZ_t) = \frac{\beta Z_{1,t}\money}{\sigma_{t}} ; \qquad
\hat{\Gamma}_0^2 = \frac{2\beta^2\kappa_1G^2}{\sigma_{t}^4} + \frac{8\beta \eta \sqrt{\kappa_1\kappa_v}\,\rho_{1v} \hat{g}_0(\vZ_t) \money}{\sigma_{t}^3}
+ \frac{8\eta^2\kappa_v \hat{g}_0(\vZ_t)^2}{\sigma_{t}^2} ;
\]
note again that $\rho_{01}$ and $\rho_{0v}$ do not enter.

\begin{figure}[h!]
\begin{tabular}{l}
(d) \scalebox{0.8}{\includegraphics*{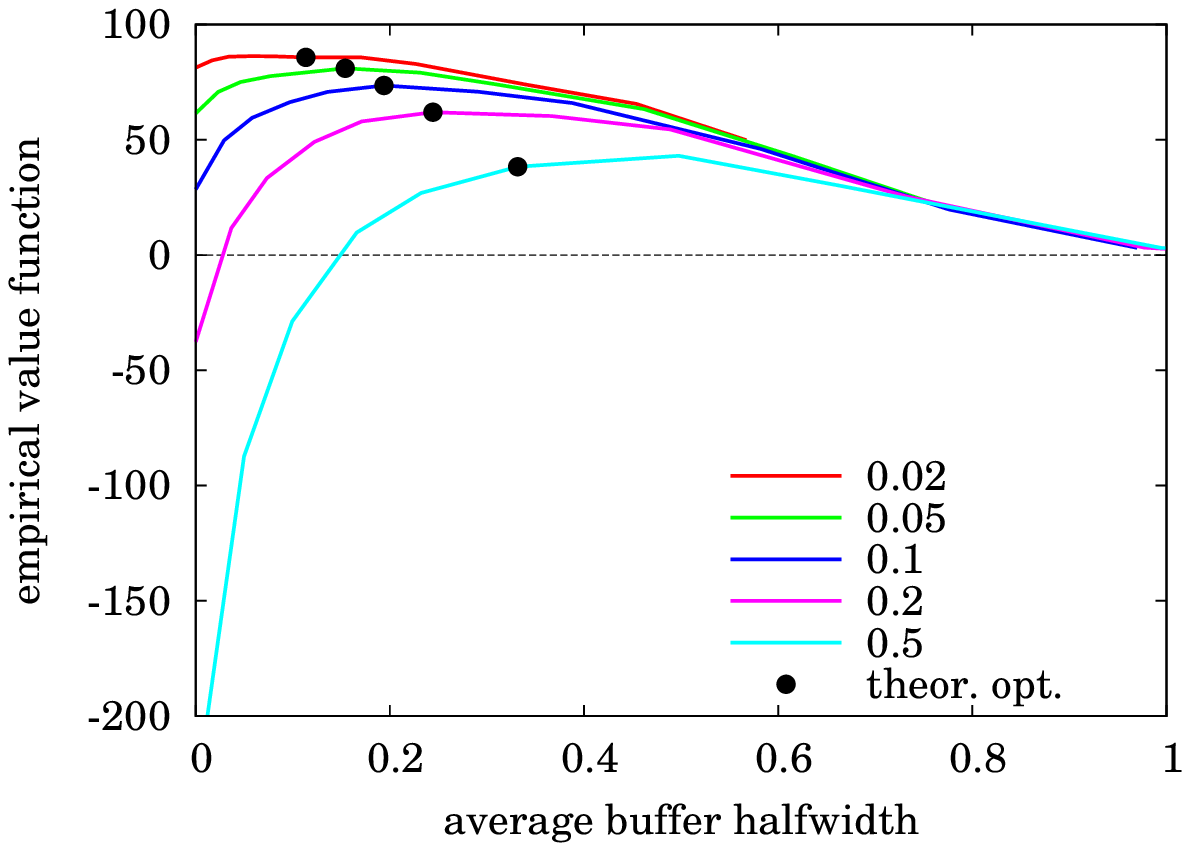}} \\
(e) \scalebox{0.8}{\includegraphics*{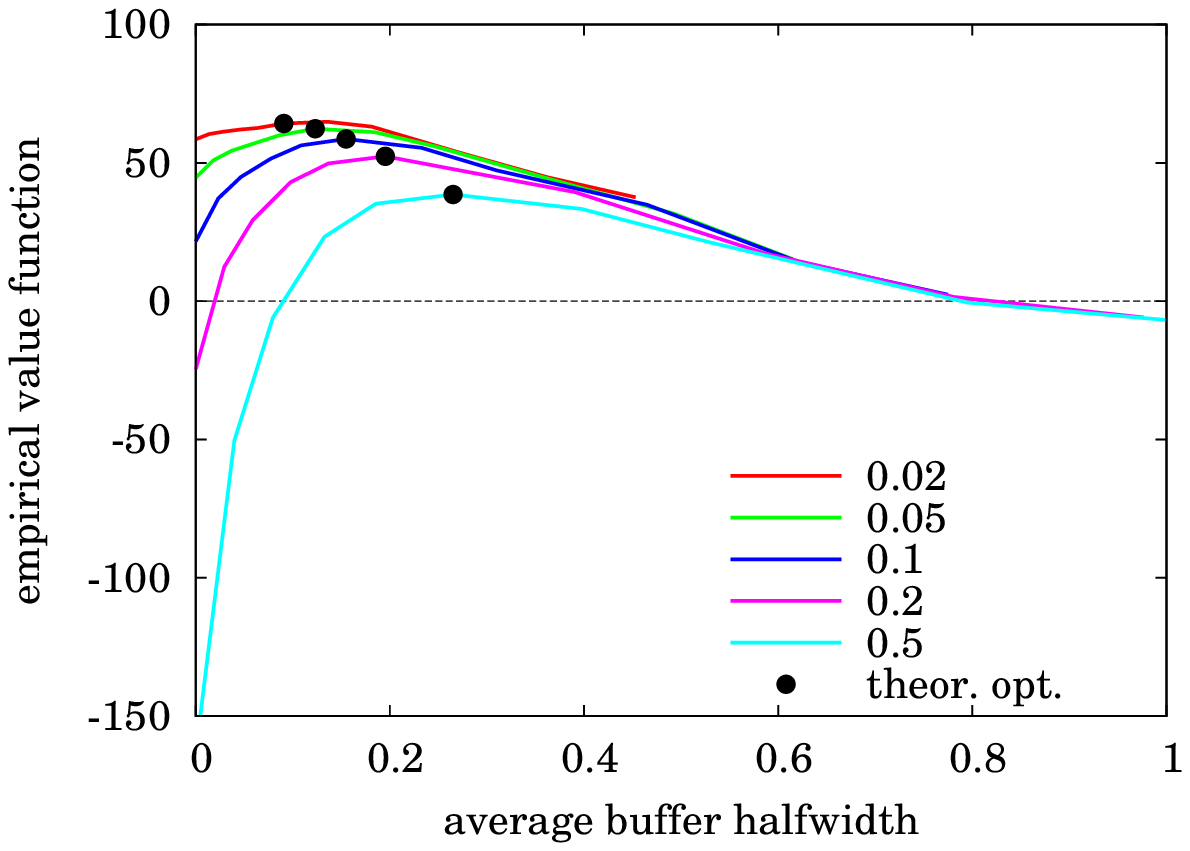}} 
\end{tabular}
\caption{\small Empirical value function vs buffer size for different models and costs.
Models: 
 (a) linear model, (b) model with nonlinear coupling, (c) linear model with stochastic volatility, (d) nonlinear coupling and stochastic volatility, (e) model with two prediction factors and using rolling estimation of $\hat{\Gamma}_0^2$. 
Cost multiplier ($\varepsilon$) as stated on graphs; $\bullet$ marks theoretical optimum (\ref{eq:soln-mf-1}).}
\label{fig:util_vs_buf}
\end{figure}

Figure~\ref{fig:util_vs_buf}(c) repeats the linear model of Figure~\ref{fig:util_vs_buf}(a), with additional parameters $\eta=0.4$, $\kappa_v=0.005$ (and $\overline{\sigma}$ stays at 0.5). One can also combine the nonlinear model of Figure~\ref{fig:util_vs_buf}(b) with stochastic volatility: see Figure~\ref{fig:util_vs_buf}(d).

\subsection*{Multiple factors}

We can easily introduce further factors alongside $Z_1$, thereby modelling multiple predictors:
\begin{eqnarray}
dX_t &=& \big(\beta_1\, \gamma_1(Z_{1,t}) +\beta_2\, \gamma_2(Z_{2,t})\big)\sigma_t \, dt + \sigma_t \, dW_{0,t} \nonumber \\
dZ_{i,t} &=& -\kappa_i Z_{i,t} \, dt + \sqrt{2 \kappa_i} \, dW_{i,t} \qquad i=1,2 \label{eq:modstochvol2}  
\end{eqnarray}
and $\sigma_t$, $Z_{v,t}$ as above.
We set $\beta_i=0.1$ and $\gamma_i(x)=\tanh(2x)$, but have the $\kappa$'s different: $\kappa_1=0.02$ and $\kappa_2=0.005$, so that the first factor reverts four times as rapidly as the second. The correlation between $dW_{1,t}$ and $dW_{2,t}$ is $\rho_{12}=0.5$; both are uncorrelated with $dW_{0,t}$ and $dW_{v,t}$.

The expression for $\hat{\Gamma}_0^2$ is now cumbersome as it depends on all the various drifts, volatilities and inter-factor correlations. We therefore use a simpler idea that we shall reuse later. It is easy enough to estimate $\hat{\Gamma}_0^2$ by forming rolling historical estimates of the quadratic variation of $\theta_t$ and of $X_t$, from the simulated time series, and taking their quotient. Using this method, we obtain Figure~\ref{fig:util_vs_buf}(e), and again the whole scheme seems to work quite well.

Consistently, the examples in Figure~\ref{fig:util_vs_buf} show that the theoretical buffering rule (\ref{eq:soln-mf-1}) generates the highest empirical value, at least by comparison with rules that are scaled up or down by some constant factor. The suggestion is that one need not plot the graphs of value vs buffer width and search for the maximum by hand, but instead trust that eq.~(\ref{eq:soln-mf-1}) is a universal law thereby saving considerable effort.

\begin{figure}[h!]
\begin{tabular}{l}
(a) \scalebox{0.8}{\includegraphics*{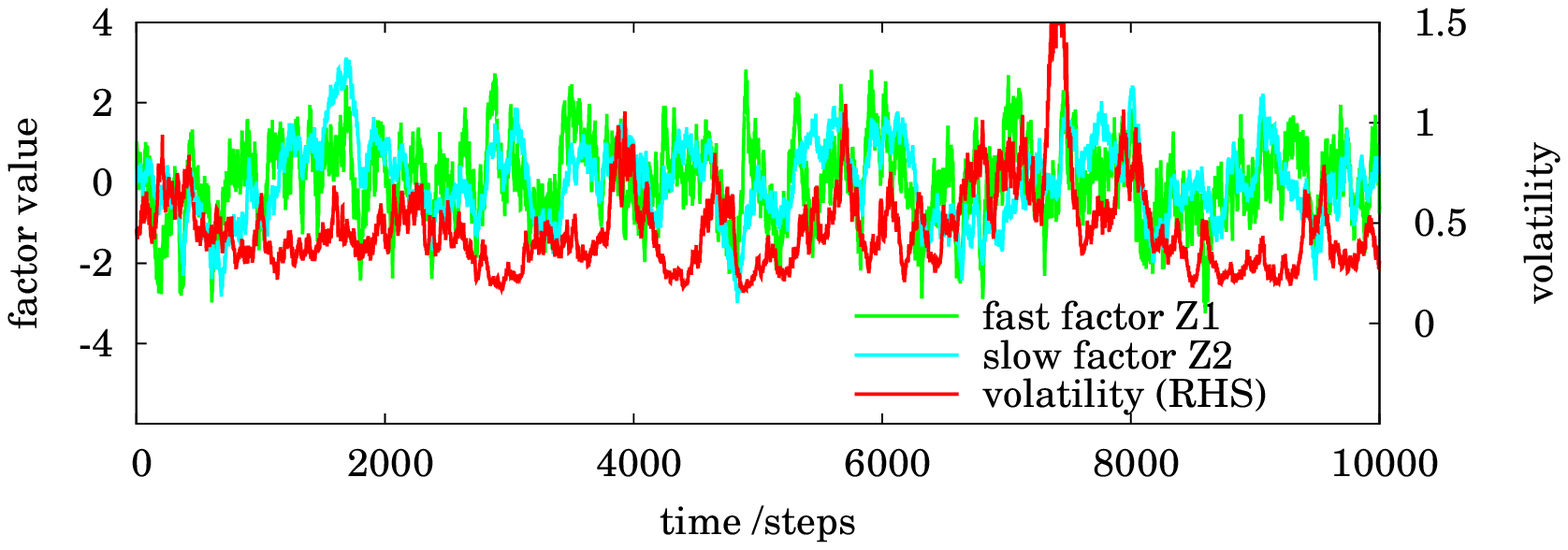}} \\
(b) \scalebox{0.8}{\includegraphics*{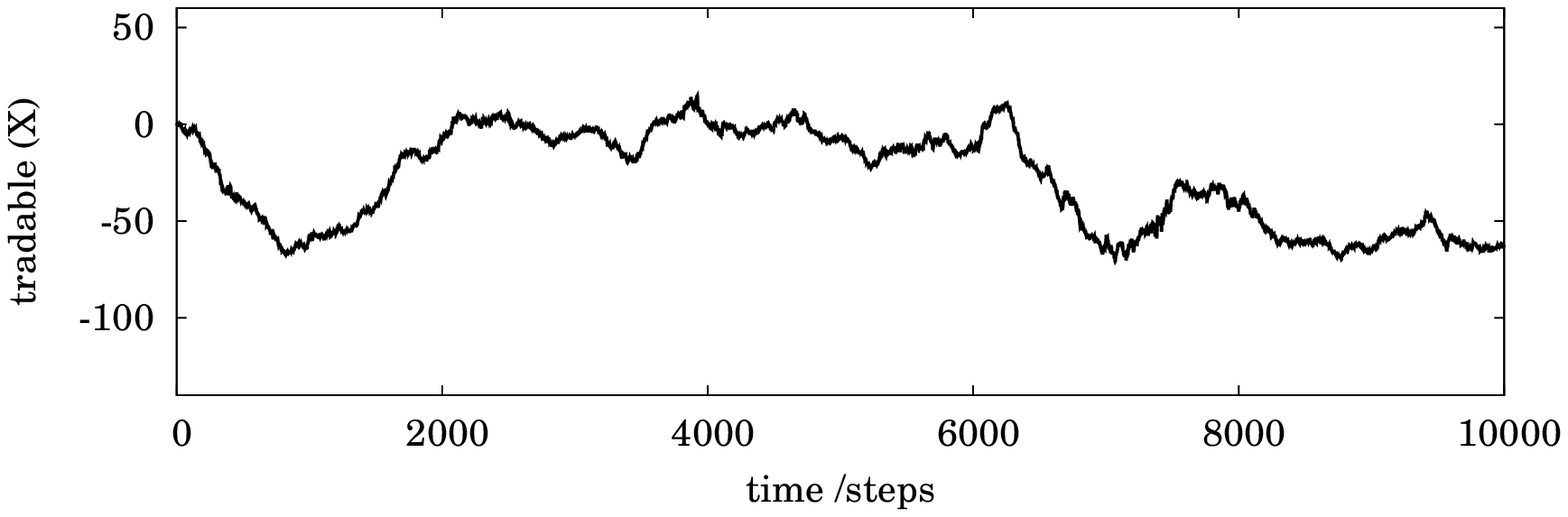}} \\
(c) \scalebox{0.8}{\includegraphics*{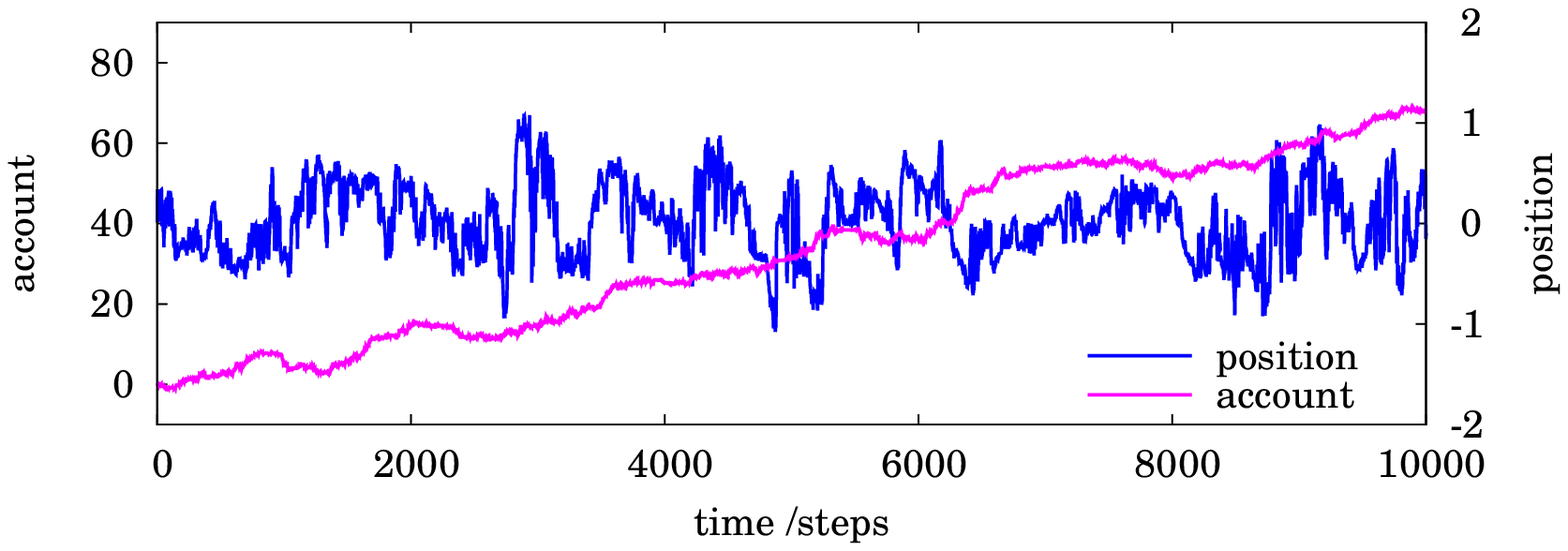}} 
\end{tabular}
\caption{\small For model with nonlinear dependence and stochastic volatility, this shows (a) the factors (return factors $Z_{1,t}$, $Z_{2,t}$ in green/cyan, and volatility $\overline{\sigma} \exp(\eta Z_{v,t}-\half \eta^2)$ in red), (b) the tradable asset $X_t$, (c) the position taken and account curve net of costs.}
\label{fig:tseries}
\end{figure}

To round off this section we give a view of the various parts of the model with nonlinear coupling and stochastic volatility. Figure~\ref{fig:tseries}(a) shows the time series of the factors $Z_{i,t}$ and the volatility multiplier $\exp(\eta Z_{v,t}-\half \eta^2)$, and Figure~\ref{fig:tseries}(b) shows the time series of the tradable $X_t$. Figure~\ref{fig:tseries}(c) shows the position $\theta_t$ when the costs are given by $\varepsilon=0.2$ and using the theoretically optimal buffer (which as can be seen from Figure~\ref{fig:util_vs_buf}(d) has average halfwidth $\approx 0.3$) and also the corresponding account curve net of costs.

\section{Examples using real data}

\subsection*{Model construction}

We return to the previously-mentioned issue of exogenous vs.\ endogenous factors. The models of the previous section lend themselves more readily to interpretation as exogenous factors, whose dynamics and interdependence are explicitly known. The models in this section are going to be of the price-driven, endogenous, type.
This is important because in constructing price-technical models one does not---and does not want to---write down a complete model for the dynamics and interdependence of the various trading signals. Rather, one identifies the signals from the time series using some recipe, and some combination of these (let us assume linear) is used to form a prediction and hence an optimal `target' position: this is completely specified by the signals and empirically-determined signal weights. For buffering (i.e.\ to get $\hat{\Gamma}_0^2$ in (\ref{eq:soln-mf-1})), one additionally needs to know the volatility of the target position. The practical solution, as anticipated, is to estimate it empirically using an historical volatility estimate from previous days' trading (in a live system) or simulated trading (in a simulated system). 


A common trending indicator is a weighted average of past returns,
\[
\mathcal{K}[X]_t = \int_{\tau=-\infty}^t K(t-\tau) \, dX_\tau.
\]
For example, the difference between two moving averages of prices, wherein the fast average exceeding the slow average is an indicator of an up-trend\footnote{A very commonly used device discussed for example in \cite[\S9]{Herbst92} and also in many online articles on technical analysis, e.g.\ {\tt www.stockcharts.com/school}}, can be written in this form.
If $X_t$ is an arithmetic Brownian motion of volatility $\sigma_X$ then the process $\mathcal{K}[X]_t$ is stationary with variance $\int_0^\infty K(\tau)^2\,d\tau \cdot \sigma_X^2$, provided the integral is convergent\footnote{For background to this result and related issues see, for example, \cite{Martin99d}.}.
This allows a normalised momentum signal to be constructed and coupled into the dynamics of the traded asset $X_t$ in the same way as in the synthetic examples,
\[
dX_t = \beta\, \gamma (Z_{t}) \sigma_{t} \, dt + \sigma_{t} \, dW_{0,t}, 
\]
thereby giving a target position $\theta_t=\beta\, \gamma(Z_{t})\money/\hat{\sigma}_{t}$.
It is necessary to estimate $\beta$ from the data either by regression or, more consistently with the approach of maximising (\ref{eq:valfn}), by directly maximising the empirical value function w.r.t.~$\beta$; the two methods are very similar.
\notthis{
However, it is possible to apply a transformation to $Z_{1,t}$ first, in the style of (\ref{eq:modlin}):
\[
dX_t = \beta\, \gamma (Z_{1,t}) \sigma_X \, dt + \sigma_X \, dW_{0,t}. 
\]
 An extreme example of this, which is often implied by articles on technical analysis (e.g.\ \cite[\S9]{Herbst92}), is to make $\gamma(x)=\mathrm{sgn}(x)$, thereby describing the binary trading rule of going long or short a fixed amount according as the momentum is positive or negative: but this is unlikely to be ideal, as a succession of large trades will be executed if the signal `wobbles' as it crosses zero, and these will incur significant cost. A compromise between the linear and binary models is to use a sigmoidal function\footnote{These are often used in control theory and in neural networks, where they are referred to as \emph{activation functions}: see e.g.~\cite{\notthis{Cybenko89,}Haykin98}.} such as $\gamma(x)=\tanh(2x)$.  This is smooth and its range is bounded, both of which are advantageous; also, it works better than either of the other two cases, but a discussion of this would take us too far from the objective of this article.
} 
One can have multiple momentum factors, of different speeds, by using $K$'s of different decay-rate, thereby giving a model like (\ref{eq:modstochvol2}).

\subsection*{Results}

\begin{figure}[h!]
\begin{center}\begin{tabular}{c}
\scalebox{0.7}{\includegraphics*{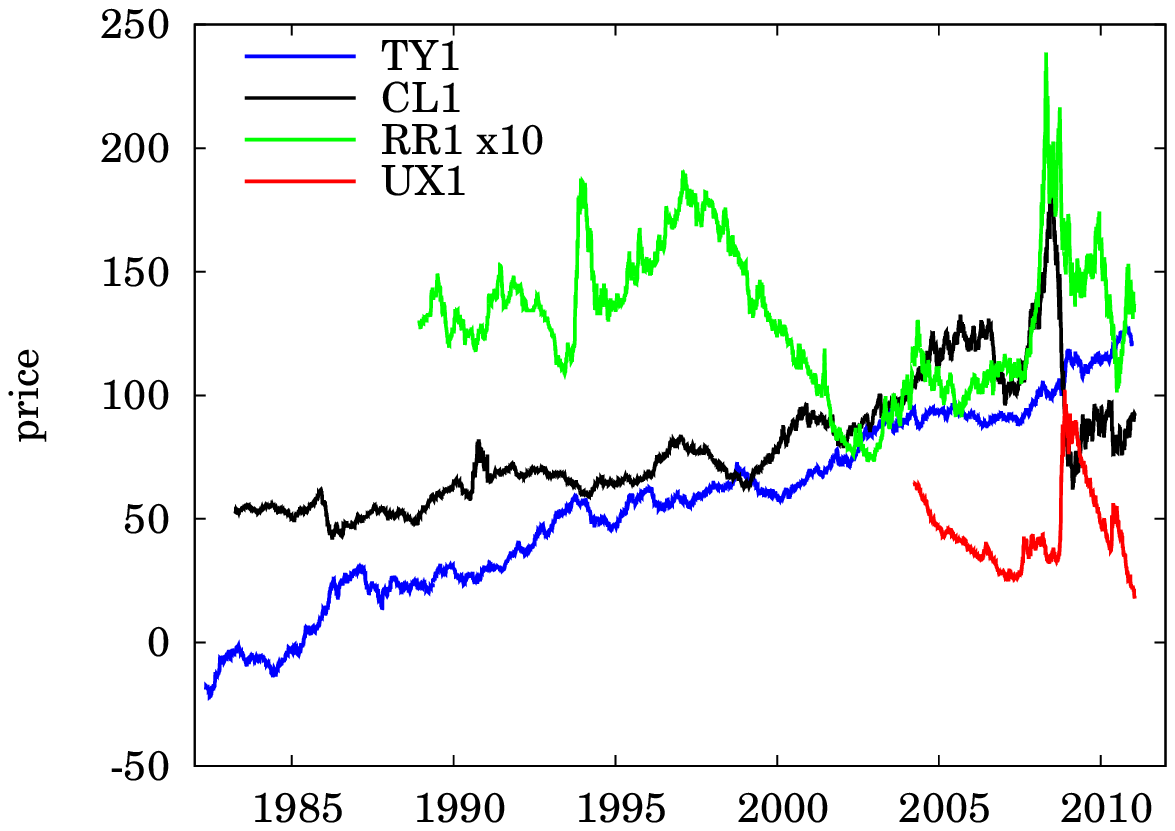}} \\
\scalebox{0.7}{\includegraphics*{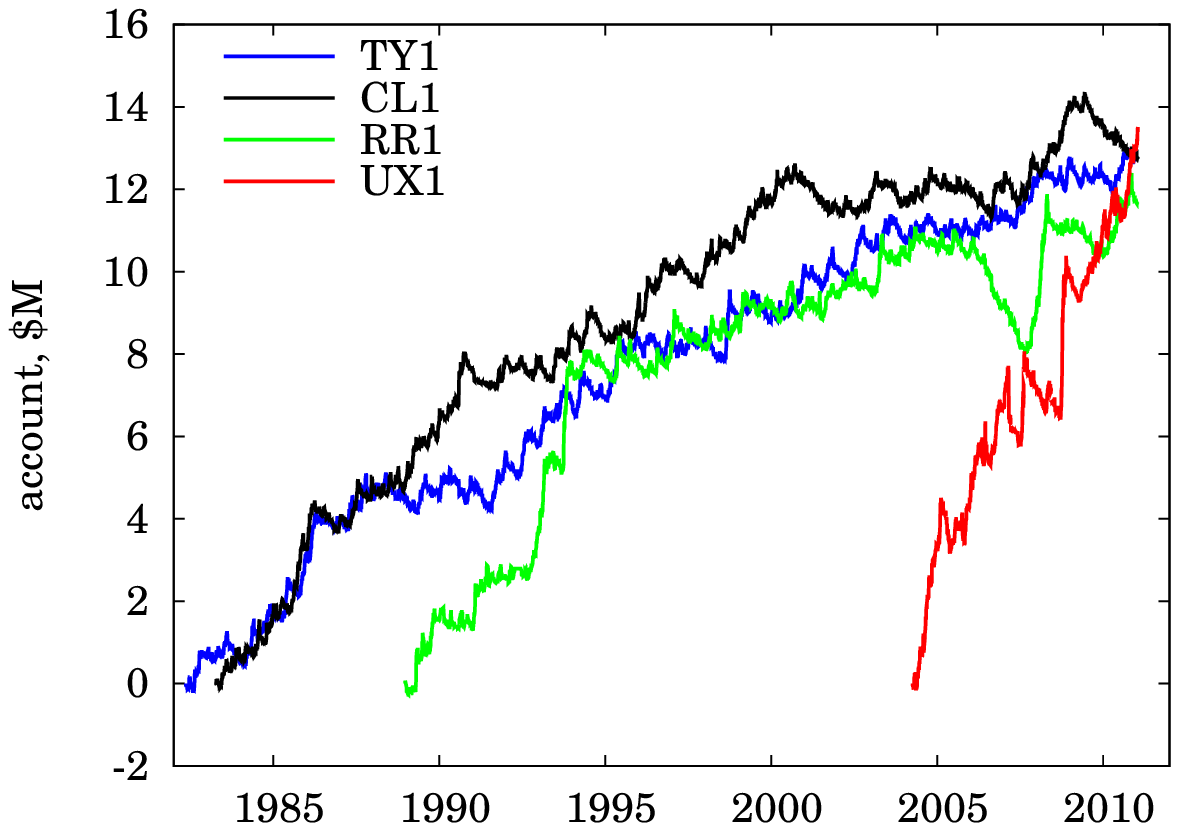}} 
\end{tabular}\end{center}
\caption{\small (Top) Time series for rolling front contracts of each of four futures, labelled with their Bloomberg tickers: US 10Y Treasury bond (TY1), crude oil (CL1), rough rice (RR1), and VIX volatility index (UX1).} (Bottom) Account curves before costs, for simple momentum strategy.
\label{fig:tseries2}
\end{figure}

We have applied the model to a variety of different futures markets, of which four are shown here from different asset classes: bonds, energies, agriculturals and the CBOE VIX contract. In each case the time series of the traded asset $X_t$ is given by stitching together the time series of the individual futures contracts, rolling 10 days before they expire\footnote{This can be done automatically in Bloomberg ({\tt GFUT <Go>}). All these contracts are designated as {\tt <Comdty>} in Bloomberg, except VIX which is {\tt UX1 <Index>}.}: see Figure~\ref{fig:tseries2}. Each of these is assumed to exhibit trending to some extent, which should result in value generation. The gearing factor of each strategy is set to $\money=\$1$M and the resulting account curves are also shown in Figure~\ref{fig:tseries2}, suggesting that the trending property is reasonably exploitable.

\begin{figure}[h!]
\begin{center}\begin{tabular}{c}
(RR1) \scalebox{0.7}{\includegraphics*{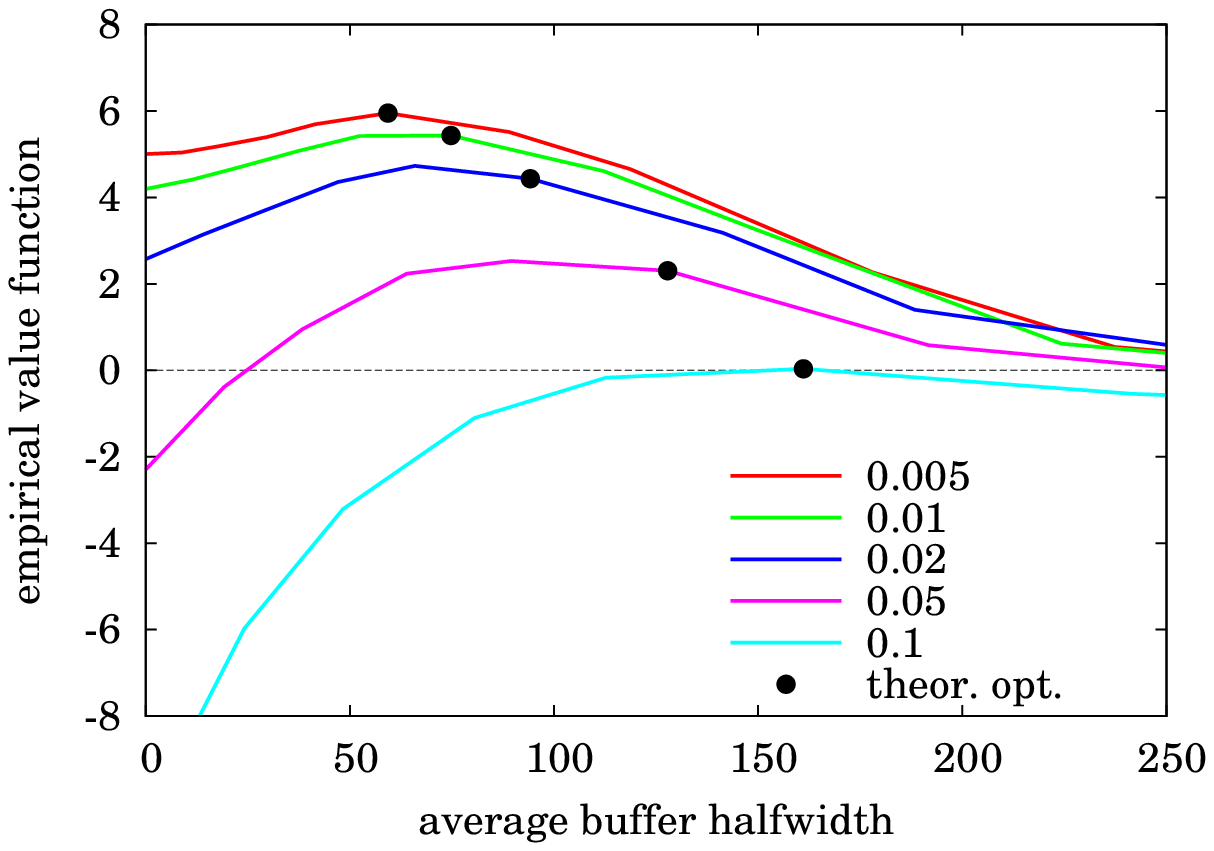}} \\
(UX1) \scalebox{0.7}{\includegraphics*{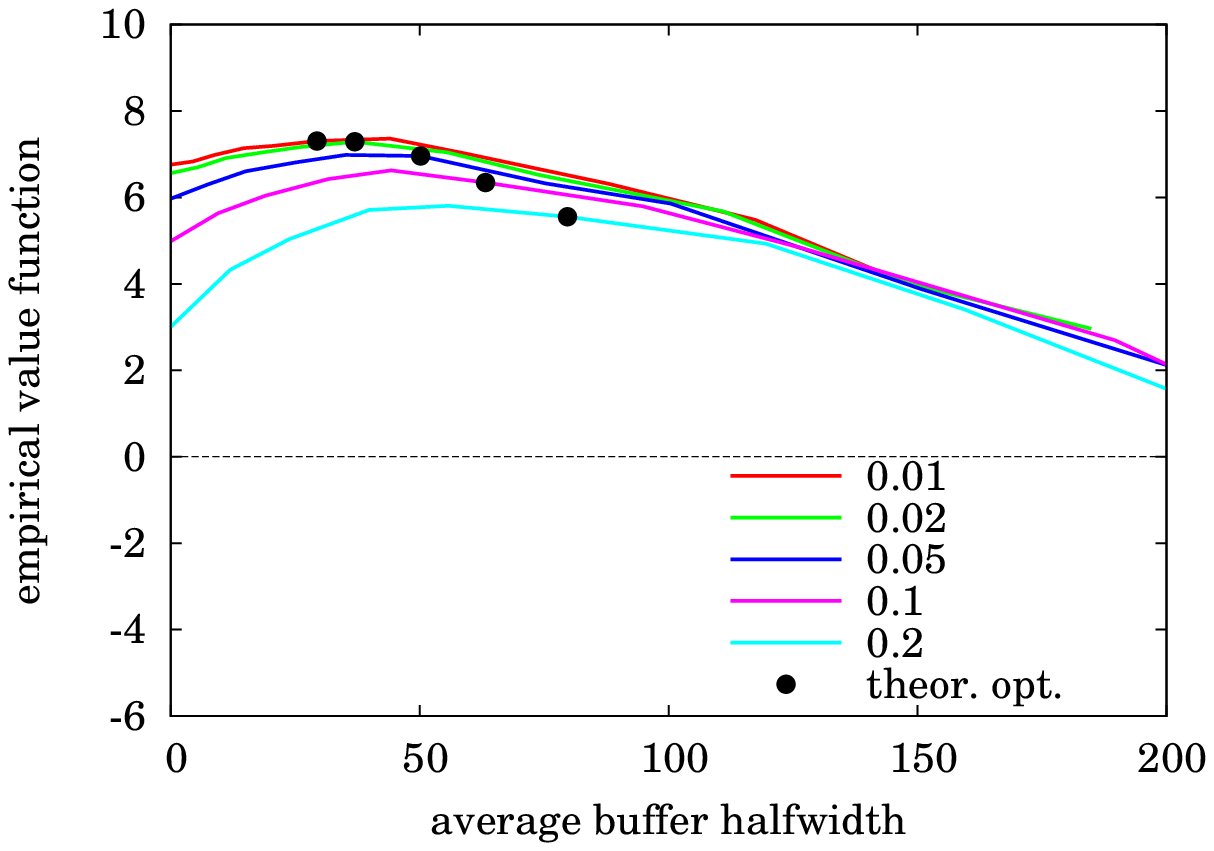}}
\end{tabular}\end{center}
\caption{\small Empirical value function vs buffer size for real examples. Cost multipliers ($\varepsilon$) are as stated on the graphs and are in contract points (not dollars). Buffer size means the number of contracts.}
\label{fig:util_vs_buf2}
\end{figure}

The effect of buffering on transaction costs is shown in Figure~\ref{fig:util_vs_buf2}. The buffer size is now a number of contracts, and the costs are in contract points so that $\varepsilon=0.005$ corresponds to a market $\frac{1}{100}$point wide, such as $89.16/89.17$. Although the curves do not have the same `ideal' shape as they do in the synthesised examples, it is reasonably clear that the theoretical optimum is reasonably optimal in practice too.

Notice for some of the contracts that trading generates almost no utility for high transaction costs (though in context, the highest costs used are much more severe than those that would be incurred in practice for these contracts). Occasionally it is seen that when the buffer is wide, the value might not quite decrease as the transaction cost increases. This is because of the hysteresis introduced by the buffer, keeping the strategy stuck in the same position for a long time (in context, years); whether it makes money or not is then a matter of chance that would be averaged out if more simulation data were available.

Notice also that for low transaction costs, the empirical value function does not go to $-\infty$ in the limit of no buffering. This is because the simulations are being done in discrete time, an issue requiring further research. 

\notthis {
Notice that the graph of empirical value function vs buffer width does not have the same `ideal' shape as it did with the synthesised examples; also for wide buffers, its value is sometimes higher for large transaction costs than for small ones. When the buffer is wide, the strategy remains stuck in the same position, i.e.\ in the NT zone, for a long time: in context, several years for the widest buffers, which is a substantial proportion of the data length.
Whether or not this position happens to make money is a matter of chance, which could be evened out by taking a longer stretch of data (if one were available). For the purposes of practical algorithm design, one is not interested in algorithms that, in effect, take a position based on where the market was months or years ago. This is because they are so path-dependent as to be unmanageable: if one simulates them from an earlier or later starting-point, the current position is likely to be completely different. The behaviour for very wide buffers is therefore of little practical importance.
} 

\section{Conclusions}

We have demonstrated a rule (\ref{eq:soln-mf-1}) for the optimal buffer, or NT, width to be applied to a diffusive factor model in the presence of proportional transaction costs and it seems to work well. For low costs\footnote{Strictly, this means lower transaction cost per unit volatility} it seems to slightly overestimate the optimal width in the `real' examples we showed, and we think this is due in part to the time discretisation in the simulation (the theory is continuous-time). We have also derived the displacement of the buffer from the costfree position, but for the reasons stated here we regard it as unimportant and we have not used it in our demonstrations.

Clearly it is important to know whether a strategy can make money after costs, even if it is profitable in theory. This is particularly true of mean-reverting or relative-value strategies, where transaction costs tend to be a higher proportion of the P\&L than for momentum strategies. Knowing how to correctly buffer a strategy is important when the transaction cost is high, as we have seen. If, despite optimising the model parameters and incorporating the buffer rule, the strategy's expected utility is still negative (which will be seen in simulation), then one knows to avoid it.

Another way of avoiding strategies that cannot reasonably work under transaction costs is to look at the typical buffer size as a proportion of the typical trading position. Once this ratio gets too large, the trading model exhibits too much hysteresis, getting stuck in the same position for perhaps months or years, and is effectively inoperable.

\subsection*{Acknowledgement}

The author thanks Torsten Sch\"oneborn, Chris Rogers and Jean-Philippe Bouchaud for helpful discussions. 

\bibliographystyle{plain}
\bibliography{../phd}



\begin{figure}[h!]
\begin{center}\begin{tabular}{c}
(a) \scalebox{0.8}{\includegraphics*{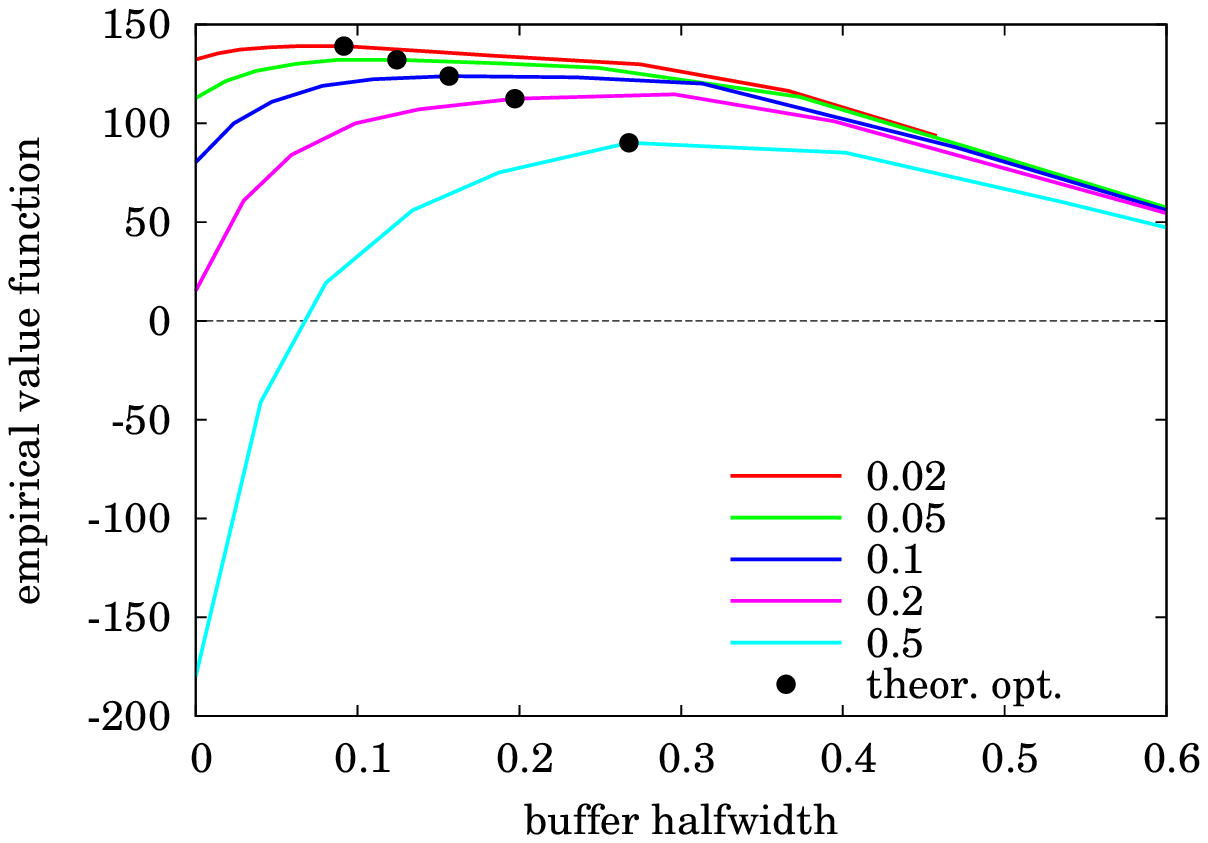}} \\
(b) \scalebox{0.8}{\includegraphics*{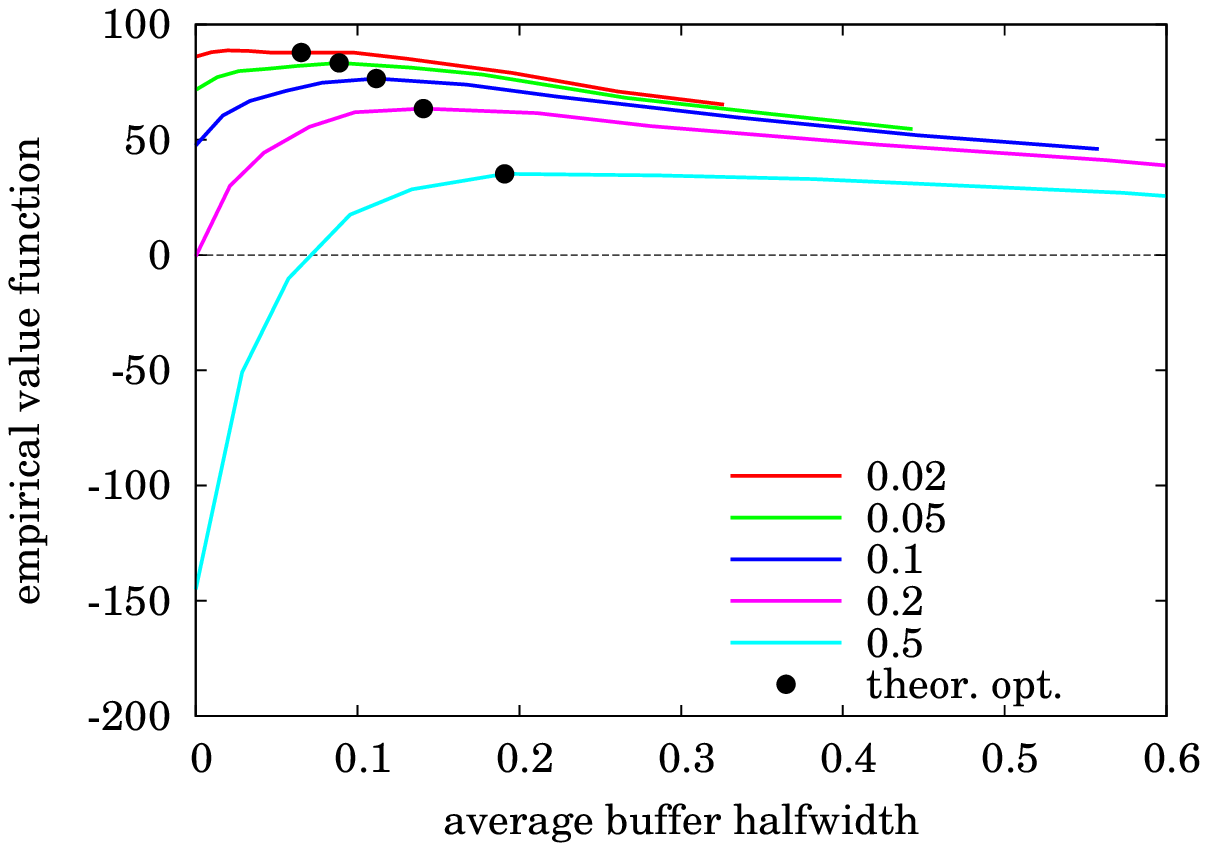}} \\
(c) \scalebox{0.8}{\includegraphics*{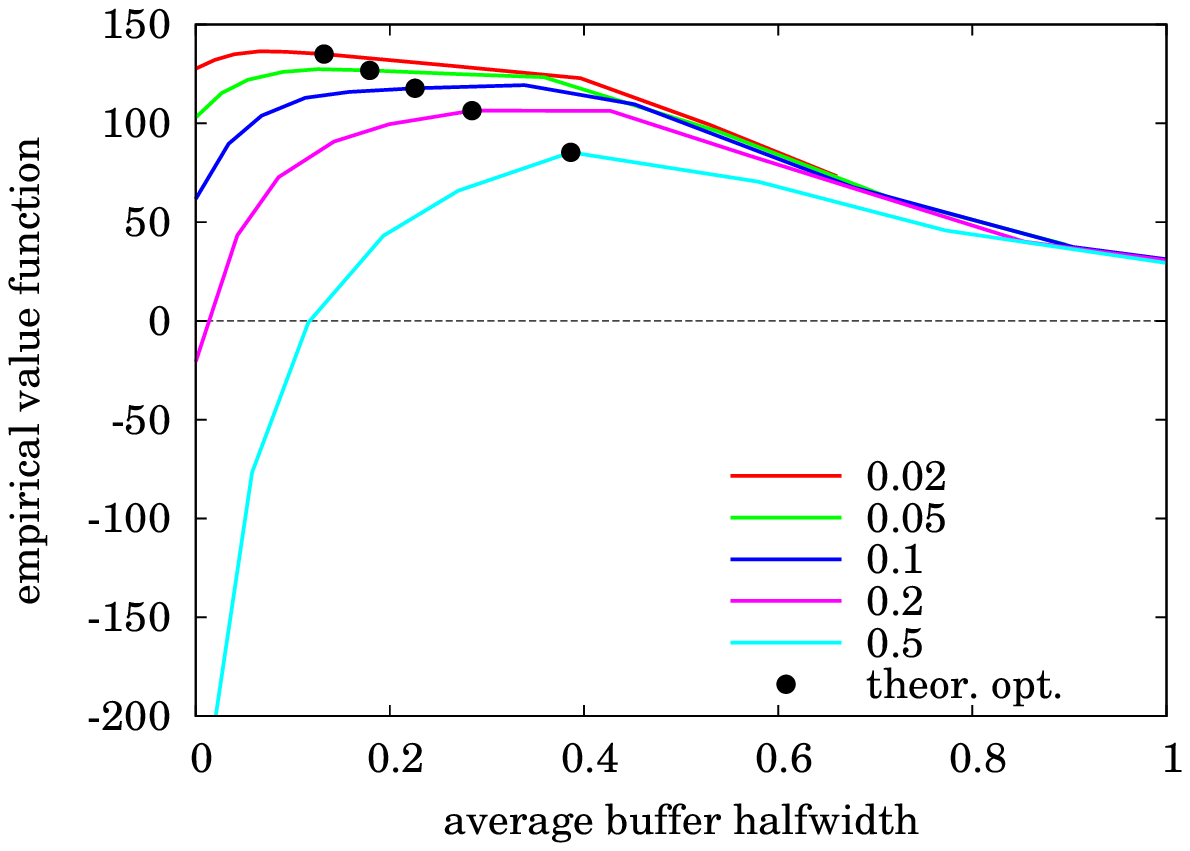}}
\end{tabular}\end{center}
\end{figure}

\begin{figure}[h!]
\begin{center}\begin{tabular}{c}
(TY1) \scalebox{0.7}{\includegraphics*{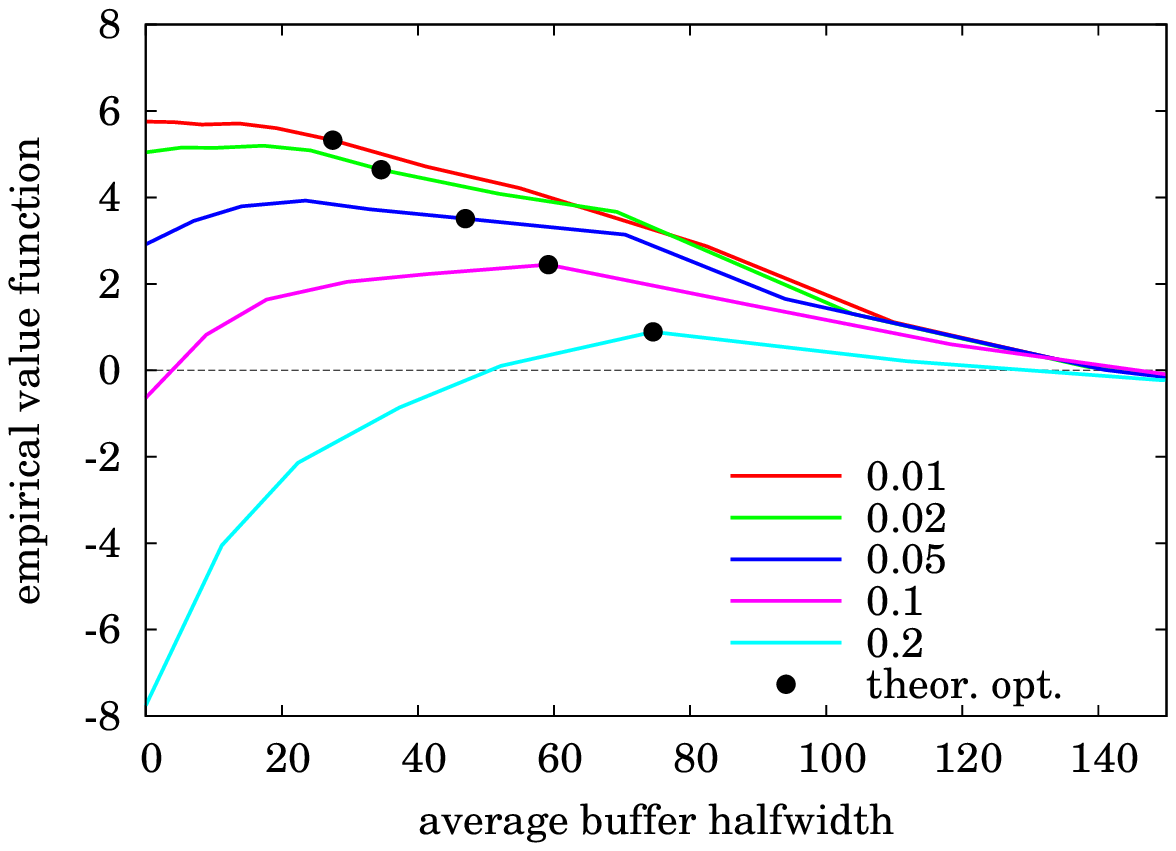}} \\
(CL1) \scalebox{0.7}{\includegraphics*{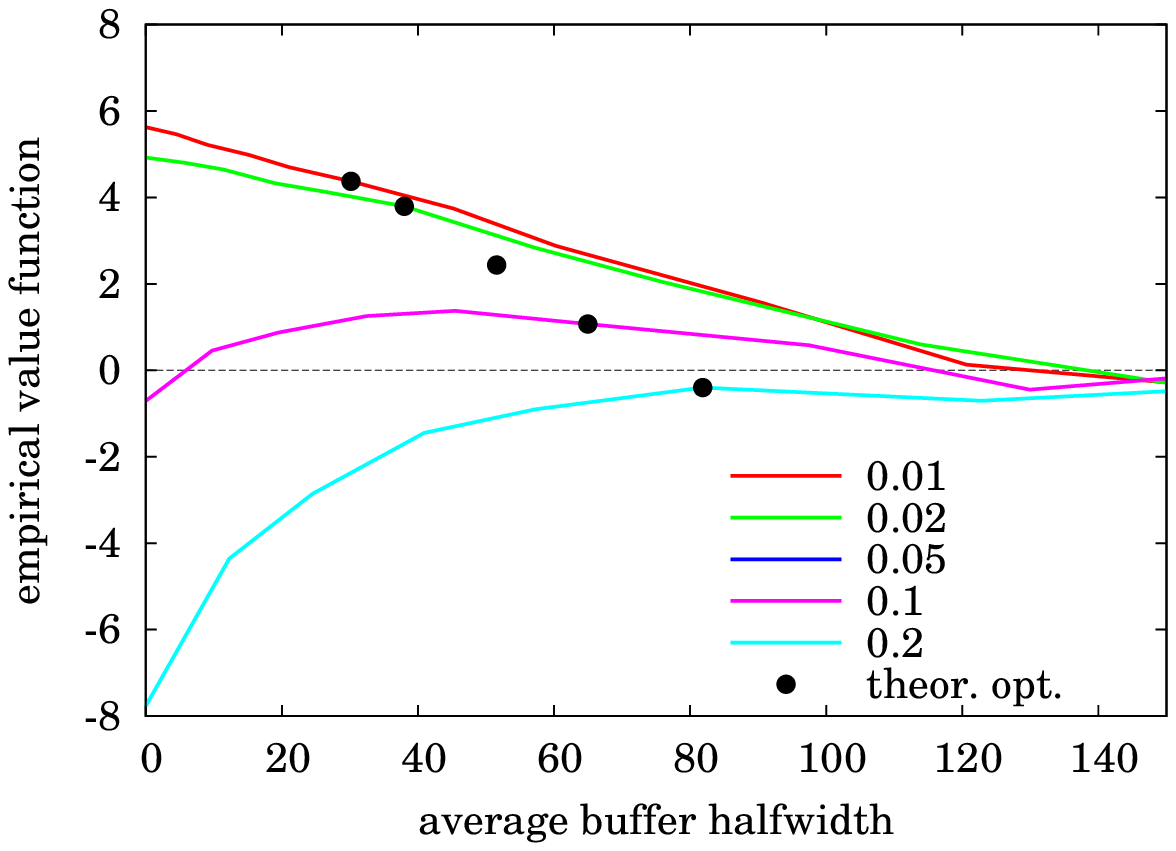}}
\end{tabular}\end{center}
\end{figure}

\clearpage 


\section{Appendix: Derivation of (\ref{eq:soln-mf-1}) and (\ref{eq:displ})}

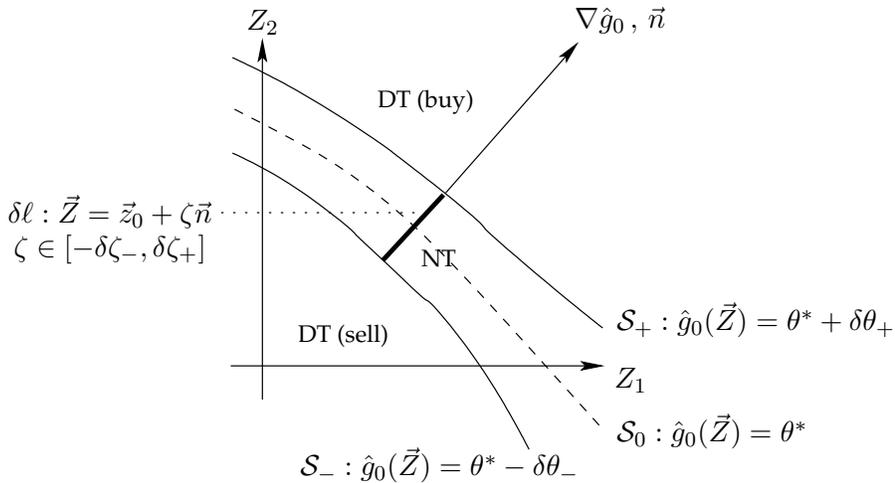
\begin{figure}[h!]
{\input{dtntdt2.pstex_t}}
\caption{\small Section through the discrete-trade and no-trade zones, for a two-dimensional factor model. All points in this plane correspond to the same position ($\theta^*$ say) in the traded asset; the dashed line is the costfree case.}
\label{fig:dtntdt2}
\end{figure}

\notthis{
The objective is to ape the derivation given in \cite{Martin11a} with specific reference to the value function in the NT zone and the analysis of small $\varepsilon$.
Although the problem (\ref{eq:model1}) ostensibly has dimension $1+\dim\vZ$, it seems possible to ignore $X_t$ as the position is not explicitly a function of it. In that case one can consider a plot of $\theta$ against $\vZ$ and then take a slice along the plane $\Pi: \theta=\theta^*$ (a constant), drawing also the `line' $\hat{g}_0(\vZ)=\theta^*$, the intersections of the NT-DT boundaries zone with $\Pi$, and the direction of most rapid increase of $g$ which is $\nabla g$. This is shown in Figure~\ref{fig:dtntdt2} for $\dim \vZ=2$, but the idea works in any number of dimensions.
} 


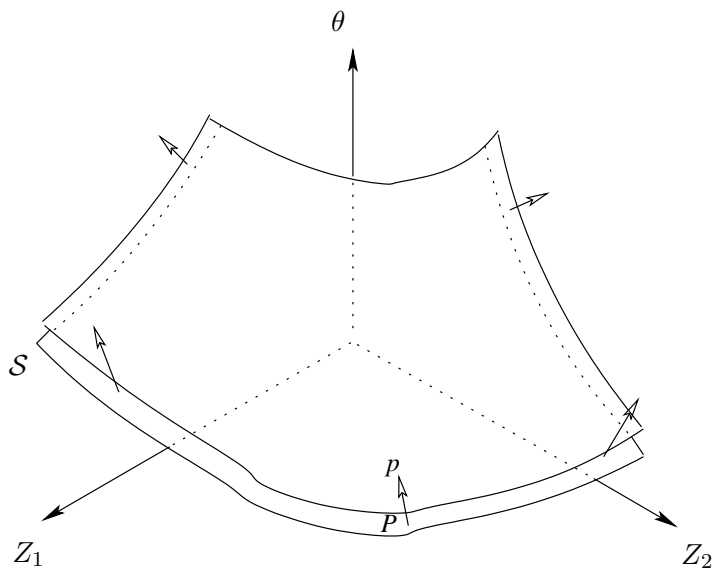
\begin{figure}[h!]
\centerline{\input{dtntdt3.pstex_t}}
\caption{\small When a smooth surface is slightly deformed, the deformation can at any point be understood as a movement in the normal direction, which typically varies from point to point (arrowed).}
\label{fig:dtntdt3}
\end{figure}

The key idea in the derivation is the reduction to one dimension, an issue that we justify now.
In the case of two factors the dependence of target position on the factors, i.e.\ $\theta=\hat{g}_0(Z_1,Z_2)$, is easily visualised in three dimensions as a smooth surface, with $\theta$ upwards (see Figure~\ref{fig:dtntdt3}; in a higher number of dimensions the derivation still works, but is harder to visualise).
The creation of the NT region involves, effectively, the creation of two copies of the surface, one placed a little above, the other a little below (with a little deformation being applied, possibly, as the spacing may not be the same everywhere).
Let $P$ be a point on a smooth surface $\mathcal{S}$, and let $\vp$ be the normal to $\mathcal{S}$ at $P$. Consider what happens when $\mathcal{S}$ is moved a small amount\footnote{Later on, in Figure~\ref{fig:dtntdt2}, we show the projection of this setup on to the $(Z_1,Z_2)$ plane. Under this projection, the vector $\vp$ becomes what is marked as $\vn$ on the diagram.}. Motions in the plane of $\mathcal{S}$, i.e.\ the two directions perpendicular to $\vp$, have no effect: it is only movement in the normal direction that does anything. The same principle holds in any number of dimensions. Thus to work out the width of the NT zone, for small costs (when it will be small), we only need to look in that normal direction, obtainable directly from $\nabla\hat{g}_0$. Consequently, all the machinery of \cite{Martin11a} can be invoked. 

Focus on the (hyper)plane $\theta=\theta^*$, pick a point $\vz$ on the optimal costfree surface\footnote{Curve when $\dim\vZ=2$ as illustrated.} $\mathcal{S}_0:\hat{g}_0(\vZ)=\theta^*$, and in the vicinity of that point let the NT zone boundaries be given locally by $\mathcal{S}_+:\hat{g}_0(\vZ)=\theta^*+\delta\theta_+$ and $\mathcal{S}_-:\hat{g}_0(\vZ)=\theta^*-\delta\theta_-$. This is shown in Figure~\ref{fig:dtntdt2} for $\dim \vZ=2$, but the construction works in any dimension.
Let $\vn$ be a unit vector in the direction $\nabla \hat{g}_0(\vz)$, which must be normal to $\mathcal{S}_0$.
Let the normal cut $\mathcal{S}_0$ at $\vZ=\vz_0$, $\mathcal{S}_+$ at $\vz_+=\vz_0+\vn\,\delta\zeta_+$ and $\mathcal{S}_-$ at $\vz_-=\vz_0-\vn\,\delta\zeta_-$, and call the line segment between these last two points $\delta \ell$. We focus on variation of the value function along $\delta \ell$.

Write $\LL$ for the infinitesimal generator of the diffusion of $\vZ$, i.e.\
\[
\LL[f] \equiv \sum_i \mu_{Z_i} \pderiv{V}{Z_i} + \half \sum_{i,j} H_{ij} \pmderiv{V}{Z_i}{Z_j}
\]
where the covariance matrix $H$ is given by $H_{ij}\,dt= \ex_t[dZ_{i,t}\,dZ_{j,t}]$.

In the NT zone there is no trading so $V_t=f(\vZ_t,\theta)$ evolves according to
\begin{equation}
(-r + \LL) f(\vZ,\theta) = -\dot{U}(\vZ,\theta)
\label{eq:valfnNT}
\end{equation}
where we have employed the usual It\^o or Feynman-Kac argument and
\[
\dot{U}(\vz,\theta) = \frac{1}{dt} \ex_t\big[\util(\theta \, dX_t) \cdl \vZ_t=\vz \big]
= \mu_X(\vz)\theta  - \frac{\sigma_X(\vz)^2\theta^2}{2\money}
\]
is the rate of accumulation of expected utility. However, we are only interested in variation along $\delta \ell$, and this is important because (\ref{eq:valfnNT}) thereby reduces to an ordinary differential equation, with $\zeta$ the coordinate in the $\vn$-direction:
\begin{equation}
(-r + \LL_{\vn}) f(\zeta,\theta) = -\dot{U}(\zeta,\theta)
\label{eq:valfnNT2}
\end{equation}
with
\begin{equation}
\LL_{\vn}[f] \equiv \Bigg( \overbrace{ \sum_i n_i\mu_{Z_i}}^{\textstyle \mu_\perp} \Bigg) \pderiv{f}{\zeta} + \half \Bigg( \overbrace{\sum_{i,j} n_iH_{ij}n_j}^{\textstyle \sigma^2_\perp} \Bigg) \pdderiv{f}{\zeta}
\label{eq:valfnNT2b}
\end{equation}
denoting the `restriction' of $\LL$ to the $\vn$-direction ($n_i$ denotes the $i$th component of $\vn$).

In the DT zone an instantaneous rebalancing is performed, so the market does not have time to move, and the value function is obtained by deducting the cost of transacting towards the NT boundary. Hence at the boundary, on the DT side, we have
\begin{equation}
\pderiv{f}{\theta}(\vz_+,\theta)  = \tcmbuy; \qquad 
\pderiv{f}{\theta}(\vz_-,\theta)  = -\tcmsell. 
\label{eq:DTbdry}
\end{equation}
We take for granted that the $\theta$-derivative is continuous at the boundary\footnote{This is discussed in the extended version of \cite{Martin11a}}.
Equations (\ref{eq:valfnNT2},\ref{eq:DTbdry}) define a two-point boundary-value problem whose solution we wish to maximise w.r.t.\ the boundary location. This is precisely the problem solved in \cite{Martin11a} with minor alterations, as follows. In (\ref{eq:valfnNT2}) there are two volatilities: the one on the LHS, labelled `$\sigma_\perp$' in (\ref{eq:valfnNT2b}), and $\sigma_X$ which occurs in $\dot{U}$ on the RHS. In \cite{Martin11a}, both are just $\sigma_X$, so we have to be a little careful. Also, if in the DT zone outside the boundary marked $\mathcal{S}_+$, at $\zeta=\delta \zeta_+$, one must \emph{buy} the asset, but in the setup of \cite{Martin11a}, in the equivalent place, one \emph{sells}. This necessitates altering a few signs.

For clarity we quickly run through the argument of \cite{Martin11a}. We first work out what is going on in the NT zone.
It is known that the equation $(-r+\LL_{\vn})f=0$ has two strictly positive solutions $f=C_+,C_-$ that are respectively increasing and decreasing functions. Let $K(\zeta,\xi)$ be the Green's function, that is, the solution to $(-r+\LL_{\vn})f=-\delta(\zeta-\xi)$; this will be positive everywhere. By standard construction of solutions to linear ODEs, the solution to (\ref{eq:valfnNT}), now written $f(\zeta,\theta)$ rather than $f(\vZ,\theta)$ as we only care about variation in the $\vn$-direction, is of the shape
\begin{equation}
f(\zeta,\theta) = \int_{-\infty}^\infty \dot{U}(\xi,\theta) K(\zeta,\xi)\, d\xi + \alpha_+(\theta)C_+(\zeta) + \alpha_-(\theta) C_-(\zeta) 
\label{eq:solnNT}
\end{equation}
i.e.\ `particular solution plus some multiple of the complementary funcion(s)'. Invoking the boundary conditions (\ref{eq:DTbdry}) we have two equations, one with $\zeta=h_+(\theta)$ at the buy boundary and one with $\zeta=h_-(\theta)$ at the sell boundary.
These give
\begin{eqnarray}
\alpha_+'(\theta) &=& \frac{I(h_-,\theta)C_-(h_+)-I(h_+,\theta)C_-(h_-) + \varepsilon_+C_-(h_-)+\varepsilon_-C_-(h_+)}{C_+(h_+)C_-(h_-)-C_+(h_-)C_-(h_+)}; \nonumber \\
\qquad
\alpha_-'(\theta) &=& \frac{I(h_+,\theta)C_+(h_-)-I(h_-,\theta)C_+(h_+) - \varepsilon_+C_+(h_-)-\varepsilon_-C_+(h_+)}{C_+(h_+)C_-(h_-)-C_+(h_-)C_-(h_+)}; \nonumber \\ &&
\label{eq:alphap}
\end{eqnarray}
where for clarity we have abbreviated $h_\pm(\theta)$ to $h_\pm$, and
\begin{equation}
I(\zeta,\theta) = \int_{-\infty}^\infty (\partial_2 \dot{U})(\xi,\theta) K(\zeta,\xi)\, d\xi; \quad
(-r+\LL_{\vn})I = -\partial_2 \dot{U}.
\label{eq:Ifn}
\end{equation}
We now wish to maximise the part of (\ref{eq:solnNT}) that is sensitive to the boundary position, and this necessitates maximising $\alpha_+(\theta)$ or $\alpha_-(\theta)$, as the first part is insensitive. As the same boundary specification must maximise the value function at all possible points in factor space (i.e.\ all $\zeta$) simultaneously, we can choose to maximise either $\alpha_+$ or $\alpha_-$ and the results should, and indeed do, give the same answer.  Furthermore, the same boundary specification must also maximise the value function at all points in position space (i.e.\ all $\theta$) simultaneously, so it is sufficient simply to maximise $\alpha_\pm'$. This means that we maximise either of the quantities in (\ref{eq:alphap}), and so differentiate w.r.t.\ $h_+$ and $h_-$. The result is a pair of coupled nonlinear equations in $h_+,h_-$. The difference between these equations gives one result, pertaining to the width of the NT zone; the sum gives a different one, pertaining to the displacement from the costfree position.
We then expand these in a Taylor series and equate terms of equal order, so that $h_+(\theta)=\zeta+\delta\zeta_+$ and $h_-(\theta)=\zeta-\delta\zeta_-$ with the $\delta$ terms small.
 Define 
\[
\invar_{i,j} = C^{(i)}_+C^{(j)}_- - C^{(i)}_-C^{(j)}_+
\]
where superscripts $^{(i,j)}$ denote derivatives; this is to be understood as a function of $\zeta$. 

For the difference, and hence the buffer width, an expression of the following form emerges\footnote{On the LHS, even powers of $\delta\zeta$ vanish by symmetry arguments, and the $\delta\zeta$ term also vanishes. Thus the algebra is laborious, as one must differentiate an already messy expression several times.}:
\[
 \varepsilon \invar_{1,0} - {\textstyle  \frac{1}{3}} \big(  
\invar_{3,1} - \invar_{3,0} \partial_1 + \invar_{1,0} \partial_1^3 \big)
I(\zeta,\theta) \cdot \delta \zeta^3 = O(\varepsilon\delta \zeta^2,\delta \zeta^5) 
\]
with $\delta \zeta = \shalf (\delta\zeta_+ + \delta\zeta_-)$. As described in \cite{Martin11a}, all expressions of the form $\invar_{i,j}/\invar_{k,l}$ relate directly to the coefficients of the ODE $(-r + \LL_{\vn})f=0$  of which $C_\pm$ are the roots (q.v., \ref{eq:valfnNT2b}): this means that one need not know what $C_\pm(\zeta)$ actually are. The second part of (\ref{eq:Ifn}) then allows further simplification.

By the Implicit Function Theorem, the optimal half-width in the $\zeta$-direction is at leading order, from \cite{Martin11a},
\[
\delta \zeta = \shalf (\delta\zeta_+ + \delta\zeta_-)  \sim \left(\frac{3\varepsilon \sigma_\perp^2}{2(\partial_1\partial_2 \dot{U})(\zeta,\theta)}\right)^{1/3}.
\]
Now $\hat{g}_0(\vZ)$ is the value of $\theta$ that maximises $\dot{U}(\vZ,\theta)$, so $(\partial_2 \dot{U})\big(\vZ,\hat{g}_0(\vZ)\big)=0$ for all $\vZ$. Differentiating,
\[
(\nabla_1 \partial_2 \dot{U})\big(\vZ,\hat{g}_0(\vZ)\big) + (\partial_2^2 \dot{U})\big(\vZ,\hat{g}_0(\vZ)\big) \nabla \hat{g}_0(\vZ) = \vec{0},
\]
with $\nabla_1$ indicating a (vector) derivative w.r.t.\ the first argument $\vZ$. Taking the scalar product with $\vn$ and noting that the derivative in the $\vn$-direction, which we denote $\partial_\zeta$, is simply $\vn\cdot\nabla$, we have
\[
(\partial_\zeta \partial_2 \dot{U})\big(\vZ,\hat{g}_0(\vZ)\big) = -\big|\nabla \hat{g}_0(\vZ)\big| (\partial_2^2 \dot{U})\big(\vZ,\hat{g}_0(\vZ)\big) 
= \big|\nabla \hat{g}_0(\vZ)\big| \sigma_X^2 \money^{-1} ,
\]
the last step following from the definition of $\dot{U}$.
Hence
\begin{equation}
\delta \zeta \sim \left(\frac{3\varepsilon \sigma_\perp^2 \money}{2 \big|\nabla \hat{g}_0(\vZ)\big| \sigma_X^2}\right)^{1/3}
\label{eq:dzeta}
\end{equation}
(this is independent of $\money$, as $\hat{g}_0(\vZ)\propto \money$).
Multiplying by $|\hat{g}_0'(\vZ)|$ gives the half-width in the $\theta$-direction:
\[
\delta \theta  \sim \left(\frac{3\varepsilon \sigma_\perp^2 \money \big|\nabla \hat{g}_0(\vZ)\big|^2 }{2 \sigma_X^2}\right)^{1/3}.
\]
But
\begin{equation}
\V_t [d\hat{g}_0(\vZ_t)] = \sum_{i,j} \big(\nabla \hat{g}_0(\vZ)\big){}_i H_{ij} \big(\nabla \hat{g}_0(\vZ)\big){}_j \, dt = \sigma_\perp^2 \big|\nabla \hat{g}_0(\vZ)\big|^2 \,dt,
\label{eq:sigmaperp}
\end{equation}
the last step by (\ref{eq:valfnNT2b}), so we find
\begin{equation}
\delta \theta  \sim \left(\frac{3\varepsilon \money \V_t [d\hat{g}_0(\vZ_t)] }{2 \V_t [dX_t]}\right)^{1/3},
\label{eq:soln-mf}
\end{equation}
as contended.
In the OU case $dX_t=-bX_t\,dt+\sigma\,dW_t$, we have $\hat{g}_0(X)=-bGX/\sigma^2$, so $\hat{\Gamma}_0^2=b^2G^2/\sigma^4$
and
\[
\delta \theta  \sim (3\varepsilon b^2 / 2 \sigma^4 )^{1/3} \money ,
\]
as previously obtained in \cite{Martin11a}.

We now turn to the displacement of the centre of the NT zone from the costfree case.  Formally we mean the following: for any $\vZ$, the optimal costfree positon is $\theta=\hat{g}_0(\vZ)$ and the boundaries of the NT zone (obtained my moving in the $\theta$-direction keeping $\vZ$ fixed) are at 
\[
\theta = \hat{g}_0(\vZ) \pm \delta \theta + \mathrm{d}\theta 
\]
where $\delta \theta$ is the halfwidth as previously obtained and we call $\mathrm{d}\theta $ the displacement.
Taylor analysis (of the sum of the nonlinear equations previously mentioned) gives an expression of the form
\[
 \varepsilon \, \delta \zeta \, \invar_{2,0} + \big(
\invar_{2,1} - \invar_{2,0} \partial_1 + \invar_{1,0} \partial_1^2  \big)
I(\zeta,\theta) \cdot \delta \zeta ^2
=  O(\varepsilon\delta \zeta^2,\delta \zeta^4) 
\]
from which (as the expression in front of $I(\zeta,\theta)$ is, up to a factor, the differential operator $(-r+\LL_{\vn})$) we find
\[
- \varepsilon \frac{\invar_{2,0}}{\invar_{1,0}} \sim 
  \frac{\delta \zeta}{ \shalf \sigma^2_\perp } (-r+\LL_{\vn}) I(\zeta,\theta).
\]
Now $(-r+\LL_{\vn})I=-\partial_2\dot{U}$, and $(\partial_2 \dot{U})\big(\vZ,\hat{g}_0(\vZ)\big)$ for all $\vZ$ by optimality, so evaluating at $\theta=\hat{g}_0(\vZ)+\mathrm{d}\theta$, i.e.\ the midpoint of the NT zone, we can approximate
\[
(\partial_2\dot{U}) \big(\vZ, \hat{g}_0(\vZ)+\mathrm{d}\theta \big) \sim (\partial^2_2 \dot{U})(\vZ,\theta) \cdot \mathrm{d}\theta = - \sigma_X^2 \money\inv \mathrm{d}\theta.
\]
Thus, as $\invar_{2,0}/\invar_{1,0}=-\mu_\perp\big/\half\sigma_\perp^2$ directly from the ODE,
\[
\varepsilon \mu_\perp \sim \delta \zeta \, \mathrm{d}\theta \, \sigma_X^2 \money\inv 
\]
and so
\[
\mathrm{d}\theta \sim \frac{\mu_\perp}{\sigma_X^2} \frac{\varepsilon \money}{\delta \zeta} .
\]
But
\[
\ex_t [d\hat{g}_0(\vZ_t)] = \sum_i \big(\nabla \hat{g}_0(\vZ)\big){}_i \mu_{Z_i}  \, dt = \mu_\perp  \big|\nabla \hat{g}_0(\vZ)\big| \,dt,
\]
so using (\ref{eq:dzeta}) for $\delta\zeta$ and recalling (\ref{eq:sigmaperp}) and the definition of $\hat{\Gamma}_0^2$ we have
\begin{equation}
\mathrm{d}\theta 
\sim
\frac{\ex_t[d\hat{g}_0(\vZ_t)]}{\V_t[dX_t]} \left(\frac{2\varepsilon^2G^2}{3\hat{\Gamma}_0^2} \right)^{1/3} .
\end{equation}
In the OU case the factor on the front works out as $-\theta b / \sigma^2$. One then has the simple result
\[
\mathrm{d}\theta \sim -\theta \cdot (2\varepsilon^2b/3\sigma^2)^{1/3},
\]
as previously obtained in \cite{Martin11a}.

\end{document}

%% file: dtntdt3.pstex_t
\begin{picture}(0,0)%
\includegraphics{dtntdt3.pstex}%
\end{picture}%
\setlength{\unitlength}{3947sp}%
\begingroup\makeatletter\ifx\SetFigFont\undefined%
\gdef\SetFigFont#1#2#3#4#5{%
  \reset@font\fontsize{#1}{#2pt}%
  \fontfamily{#3}\fontseries{#4}\fontshape{#5}%
  \selectfont}%
\fi\endgroup%
\begin{picture}(4648,3557)(1439,-4355)
\put(1464,-4286){\makebox(0,0)[lb]{\smash{{\SetFigFont{11}{13.2}{\rmdefault}{\mddefault}{\updefault}$Z_1$}}}}
\put(5664,-4299){\makebox(0,0)[lb]{\smash{{\SetFigFont{11}{13.2}{\rmdefault}{\mddefault}{\updefault}$Z_2$}}}}
\put(3464,-949){\makebox(0,0)[lb]{\smash{{\SetFigFont{11}{13.2}{\rmdefault}{\mddefault}{\updefault}$\theta$}}}}
\put(1439,-3111){\makebox(0,0)[lb]{\smash{{\SetFigFont{11}{13.2}{\rmdefault}{\mddefault}{\updefault}$\mathcal{S}$}}}}
\end{picture}%

%% file: dtntdt2.pstex_t
\begin{picture}(0,0)%
\includegraphics{dtntdt2.pstex}%
\end{picture}%
\setlength{\unitlength}{3947sp}%
\begingroup\makeatletter\ifx\SetFigFont\undefined%
\gdef\SetFigFont#1#2#3#4#5{%
  \reset@font\fontsize{#1}{#2pt}%
  \fontfamily{#3}\fontseries{#4}\fontshape{#5}%
  \selectfont}%
\fi\endgroup%
\begin{picture}(4235,3019)(739,-2392)
\put(4551,-1786){\makebox(0,0)[lb]{\smash{{\SetFigFont{11}{13.2}{\rmdefault}{\mddefault}{\updefault}$Z_1$}}}}
\put(2239,464){\makebox(0,0)[lb]{\smash{{\SetFigFont{11}{13.2}{\rmdefault}{\mddefault}{\updefault}$Z_2$}}}}
\put(2576,-2336){\makebox(0,0)[lb]{\smash{{\SetFigFont{11}{13.2}{\rmdefault}{\mddefault}{\updefault}$\mathcal{S}_-:\hat{g}_0(\vZ)=\theta^*-\delta\theta_-$}}}}
\put(4564,-1424){\makebox(0,0)[lb]{\smash{{\SetFigFont{11}{13.2}{\rmdefault}{\mddefault}{\updefault}$\mathcal{S}_+:\hat{g}_0(\vZ)=\theta^*+\delta\theta_+$}}}}
\put(4564,-2124){\makebox(0,0)[lb]{\smash{{\SetFigFont{11}{13.2}{\rmdefault}{\mddefault}{\updefault}$\mathcal{S}_0:\hat{g}_0(\vZ)=\theta^*$}}}}
\put(4301,476){\makebox(0,0)[lb]{\smash{{\SetFigFont{11}{13.2}{\rmdefault}{\mddefault}{\updefault}$\nabla\hat{g}_0\, , \,\vn$}}}}
\put(739,-749){\makebox(0,0)[lb]{\smash{{\SetFigFont{11}{13.2}{\rmdefault}{\mddefault}{\updefault}$\delta\ell : \vZ=\vz_0+\zeta \vn$}}}}
\put(776,-961){\makebox(0,0)[lb]{\smash{{\SetFigFont{11}{13.2}{\rmdefault}{\mddefault}{\updefault}$\zeta\in[-\delta\zeta_-,\delta\zeta_+]$}}}}
\end{picture}%

%% file: arxiv_mftrcost.bbl
\begin{thebibliography}{10}

\bibitem{Bjork98}
T.~Bj\"ork.
\newblock {\em Arbitrage Theory in Continuous Time}.
\newblock OUP, 1998.

\bibitem{Davis90}
M.~H.~A. Davis and A.~R. Norman.
\newblock Portfolio selection with transaction costs.
\newblock {\em Math. Oper. Research}, 15(4):676--713, 1990.

\bibitem{Garleanu09}
N.~Garleanu and L.~H. Pedersen.
\newblock Dynamic trading with predictable returns and transaction costs, 2009.
\newblock NBER Working paper, Berkeley Univ, {\tt
  www.nber.org/papers/w15205.pdf}.

\bibitem{Haykin98}
S.~Haykin.
\newblock {\em Neural Networks: A Comprehensive Foundation}.
\newblock Prentice Hall, 1998.

\bibitem{Herbst92}
A.~F. Herbst.
\newblock {\em Analyzing and Forecasting Futures Prices}.
\newblock Wiley, 1992.

\bibitem{Kallsen99}
J.~Kallsen.
\newblock A utility maximisation approach to hedging in incomplete markets.
\newblock {\em Math. Meth. Oper. Res.}, 50(2):321--338, 1999.

\bibitem{Lataillade12}
{J. de} Lataillade, C.~Deremble, M.~Potters, and J.-P. Bouchaud.
\newblock Optimal trading with linear costs.
\newblock {\em {\tt arXiv:1203.5957}}, 2012.

\bibitem{Martin99d}
R.~J. Martin.
\newblock Autoregression and irregular sampling: Spectral estimation.
\newblock {\em Signal Processing}, 77:139--157, 1999.

\bibitem{Martin11a}
R.~J. Martin and T.~Sch\"oneborn.
\newblock Mean reversion pays, but costs.
\newblock {\em RISK}, 24(2):84--89, 2011.
\newblock Full vsn at {\tt arxiv.org/pdf/1103.4934}.

\bibitem{Rogers04}
L.~C.~G. Rogers.
\newblock Why is the effect of proportional transaction costs
  ${O}(\delta^{2/3})$?
\newblock In {\em Mathematics of Finance, AMS Contemporary Maths. Series 351},
  pages 303--308. AMS, 2004.
\newblock Also covered by S.~Shreve in plenary talk at SIAM Annual Meeting,
  July 16, 2010; slides at {\tt www.math.cmu.edu/users/shreve}.

\bibitem{Shreve94}
S.~E. Shreve and H.~M. Soner.
\newblock Optimal investment and consumption with transaction costs.
\newblock {\em Ann. Appl. Prob.}, 4(3):609--692, 1994.

\bibitem{Whalley97}
A.~E. Whalley and P.~Wilmott.
\newblock An asymptotic analysis of an optimal hedging model for option pricing
  with transaction costs.
\newblock {\em Math. Fin.}, 7(3):307--324, 1997.

\bibitem{Zakamouline06}
V.~I. Zakamouline.
\newblock European option pricing and hedging with both fixed and proportional
  transaction costs.
\newblock {\em J. Econ. Dyn. \& Control}, 30:1--25, 2006.

\end{thebibliography}
